\documentclass[square, comma, sort&compress,12pt]{elsarticle}

\usepackage{amssymb}
\usepackage{amsthm}
\usepackage{framed} 
\usepackage{amsmath,color}
\usepackage{mathrsfs}
\usepackage{graphicx}
\usepackage{epstopdf}
\usepackage{float}
\usepackage{caption}
\usepackage{subcaption}
\usepackage{bm}
\usepackage{bbm}
\usepackage{mathrsfs}
\usepackage{cleveref}
\usepackage{soul}
\usepackage{accents}
\usepackage{color,soul} 
\usepackage{color} 
\usepackage{nomencl} 
\soulregister\citep7 
\soulregister\citet7 
\soulregister\citealp7 
\newsavebox{\measurebox} 
\usepackage{titlesec} 
\usepackage{tabu} 
\usepackage{longtable}
\usepackage{multirow} 
\usepackage{graphicx}
\usepackage[table,xcdraw]{xcolor}
\biboptions{numbers,sort&compress} 
\usepackage{enumitem}
\usepackage[T1]{fontenc}
\usepackage{lmodern}

\usepackage[margin=1.7cm]{geometry}

\usepackage{lineno}
\journal{Materials \& Design}

\makeatletter
\def\@author#1{\g@addto@macro\elsauthors{\normalsize%
    \def\baselinestretch{1}%
    \upshape\authorsep#1\unskip\textsuperscript{%
      \ifx\@fnmark\@empty\else\unskip\sep\@fnmark\let\sep=,\fi
      \ifx\@corref\@empty\else\unskip\sep\@corref\let\sep=,\fi
      }%
    \def\authorsep{\unskip,\space}%
    \global\let\@fnmark\@empty
    \global\let\@corref\@empty  
    \global\let\sep\@empty}%
    \@eadauthor={#1}
}
\makeatother

\titleformat{\paragraph}
{\normalfont\normalsize\itshape}{\theparagraph}{1em}{}
\titlespacing*{\paragraph}
{0pt}{3.25ex plus 1ex minus .2ex}{1.5ex plus .2ex}

\begin{document}

\begin{frontmatter}


\title{Process parameter sensitivity of the energy absorbing properties of additively manufactured metallic cellular materials}


\author{M. Simoes\corref{cor1}\fnref{Cav}}
\ead{ms2452@cam.ac.uk}

\author{J.A. Harris\fnref{Cam}}

\author{S. Ghouse\fnref{Imp}}

\author{P.A. Hooper\fnref{Imp}}

\author{G.J. McShane\fnref{Cam}}

\address[Cav]{Department of Physics, Cambridge University, CB3 0HE Cambridge, UK}

\address[Cam]{Department of Engineering, Cambridge University, CB2 1PZ Cambridge, UK}

\address[Imp]{Department of Mechanical Engineering, Imperial College London, SW7 2AZ London}

\cortext[cor1]{Corresponding author.}

\begin{abstract}	

\noindent Additive Manufacturing (AM) has enabled the fabrication of metallic cellular materials that are of interest in the design of lightweight impact resistant structures. However, there is a need to understand the interactions between: (i) the material architecture, (ii) the AM process parameters, and (iii) the as-built geometry, microstructure and energy absorbing properties. In this work, we investigate the quasi-static and dynamic behaviour of cellular materials manufactured from 316L stainless steel using laser powder bed fusion (LPBF). Four cellular architectures are considered (octet lattice, lattice-walled square honeycomb, origami and square honeycomb), as well as three sets of AM process parameters, characterised by laser powers of 50, 125 and 200 W. The exposure time is adjusted to deliver the same total heat input. The 125 W case leads to material with the highest strength and ductility. The cellular materials with this process variant match their nominal densities most closely, and have the highest strength and energy absorption. Either reducing (50 W) or increasing (200 W) the power leads to a significant increase in porosity, reducing strength and energy absorption. However, we find that changes due to process-induced porosity have a smaller influence than those resulting from the choice of cellular architecture. 

\end{abstract}


\end{frontmatter}

	

\section{Introduction}
\label{Sec:Introduction}

High performance energy absorbing materials are of interest in a wide range of industry sectors, such as defence, aerospace, rail and automotive. There has been extensive research in recent years into the properties, design and manufacture of lightweight metallic cellular materials \cite{Sombatmai2021,Soro2022,Lei2022}. These materials can achieve high specific energy absorption, and span a wide property space via the choice of relative density, cellular architecture (e.g. strut connectivity and orientation) and cell wall material \cite{Gibson1997,Fleck2010}. Examples include metallic foams \cite{Ashby2000}, pyramidal and tetragonal lattices \cite{Deshpande2001c,Radford2006}, honeycombs \cite{Cote2004} and egg-box geometries \cite{Deshpande2003}. Understanding the relationship between cellular architecture and compressive strength at high strain rates is essential for the design of energy absorbing materials for applications such as blast and impact protection. Important considerations include material strain rate effects and inertial stabilisation of cell buckling \cite{Calladine1984}, which in turn may be influenced by the cellular architecture and processing history.

More recently, developments in additive manufacturing (AM) have enabled metallic cellular materials to be fabricated with geometries, or alloy-geometry combinations, that would be impractical or impossible using established manufacturing techniques \cite{Blakey-Milner2021}. Further benefits of the AM route include consolidating complex assemblies into single parts, and avoiding the cost and waste of subtractive manufacture of such assemblies \cite{Rehme2006, Shen2010}. Furthermore, the complex thermal history present in laser-based AM techniques leads to a non-equilibrium solidification process which may result in beneficial microstructural changes. For example, high cooling rates (e.g., $10^3$-$10^8$ Ks$^{-1}$) lead to the formation of fine microstructures, which provide mechanical property benefits \cite{TaheriAndani2016}.

The focus of the current investigation is the energy absorbing performance of additively manufactured metallic cellular materials. A number of investigations have considered the compressive response of AM lattice structures, i.e. networks of bars \cite{McKown2008, Shen2010a, Ushijima2011}. To date, more data has been generated on the quasi-static properties, with relatively little on the dynamic compressive response, see Ref. \cite{Medvedev2022} for a comprehensive review. Smith et al. \cite{Smith2011} studied the performance of AM 316L stainless steel lattices for impact protection. Using quasi-static compression tests, and impact tests up to 32 m/s, Smith et al. \cite{Smith2011} report that AM was advantageous in producing lattice structures that offer impact protection due to their uniform collapse response. McKown et al. \cite{McKown2008} investigated experimentally the quasi-static and dynamic compressive response of steel AM lattice structures, providing data at high strain rates. Energy absorption values between 3 and 6 kJ/kg were reported by Smith et al. and McKown et al. \cite{McKown2008, Smith2011}, measured up to 50\% compressive strain. Ozdemir et al. \cite{Ozdemir2015} presented the energy absorption behaviour and failure modes of AM Ti6Al4V cubic and diamond lattice structures produced using electron beam melting (EBM). Both quasi-static compression and dynamic loading using the Hopkinson pressure bar were considered. Tancogne-Dejean et al. \cite{Tancogne-Dejean2016} investigated the dynamic behaviour of octet truss micro-lattices, revealing a significant effect of the strut geometry and the relative density. Chang et al. \cite{Chang2021} investigated the dynamic behaviour of superimposed 316L lattice structures, exhibiting a superior performance. Triply periodic minimal surface (TPMS) topologies have also received significant attention \cite{AlMahri2021,Li2021c,Saremian2021}, showing energy absorption values within the 30-40 J/kg range.

Higher specific energy absorption levels have been reported for facet-based metallic cellular materials, such as the square honeycomb. Harris and McShane \cite{Harris2021} compared the energy absorbing performance of conventionally manufactured (i.e.\ brazed sheet) stainless steel square honeycombs with those produced using laser powder bed fusion (LPBF). Up to 50\% compressive strain, energy absorption values up to about 25 kJ/kg have been reported for traditionally manufactured honeycombs \cite{Cote2004, Radford2006}, and 35 kJ/kg for LPBF square honeycombs \cite{Harris2017,Harris2020}. The LPBF process leads to a high material yield strength, due to the fine microstructure. The LPBF square honeycombs also tend to have a higher relative density, due to process limitations on the minimum cell wall thickness, which can reduce post-buckling softening. Considering alternative facet-based geometries, Harris and McShane \cite{Harris2020} considered a LPBF origami-inspired cellular material (the stacked Miura-ori). It has lower specific energy absorption compared to the square honeycomb, due to the bending-dominated architecture, albeit with more scope for property tuning, grading and further performance optimisation. Due to the complex pattern of the facets, the stacked Miura-ori architecture is particularly suited to an AM fabrication route. 

These results show the potential of AM cellular materials for energy absorption. However, the AM process parameters have to be carefully controlled to minimise defects \cite{Ioannidou2022}. In particular, the presence of porosity is an issue that impacts the mechanical properties of metallic AM parts. For powder-based processes such as LPBF, controlling this requires good powder preparation, optimisation of the laser parameters, and control of the conditions within the fabrication chamber \cite{TaheriAndani2016}. Another quality issue is variation between the nominal geometry and the as-fabricated AM component \cite{TaheriAndani2016}. Nonetheless, optimisation of the processing parameters can minimise geometry and property variations, leading to material with high density and mechanical strength \cite{TaheriAndani2016, Shen2010}. Sames et al. \cite{Sames2016a} reported that with optimised AM parameters, process-induced porosity can be reduced to less than 1\% in direct energy deposition (DED), selective laser melting (SLM) and electron beam melting (EBM) AM techniques.

For AM metallic cellular materials, the architecture - the spatial distribution of solid material - has the potential to influence microstructure development during AM processing. Previous investigations using LPBF have focussed either on one set of AM process parameters applied to a range of different cellular geometries, or on the influence of different process parameters (in particular laser power and exposure time) applied to the same cellular geometry \cite{Brooks2005, McKown2008, Shen2010, Shen2010a, Smith2011, Ozdemir2015}. Typically, these focus on lattice structures (i.e.\ truss-like arrays of bars), and in particular the BCC unit-cell configuration. LPBF facet-based architectures (i.e. arrays of plates) have been investigated by Harris et al. \cite{Harris2017, Harris2020, Harris2021}. However, the LPBF process parameters were not varied in those experiments.

The aim of the current investigation is to identify the relationship between cellular architecture, variations in AM process parameters (and the consequent variations in microstructure and material defects), and the energy absorption performance for a wide range of AM cellular materials. Laser powder bed fusion (LPBF) is used to fabricate the metallic cellular materials from 316L stainless steel. This alloy is chosen for its amenability to the LPBF process, so as to minimise additional alloy-related effects. For each cellular geometry, the LPBF processing parameters are varied to modify the powder melting step. Specifically, the laser power and the exposure time are varied, while maintaining a fixed total heat input, as described subsequently. Three process parameter sets/variants are considered for each type of cellular material. Four contrasting cellular material geometries are investigated here, with each subjected to identical LPBF process variations. Two are strut-based lattice structures: the octet truss \cite{Deshpande2001} and the lattice-walled square honeycomb \cite{Harris2017}. Two are facet-based: the origami-inspired stacked Miura-ori structure \cite{Harris2020,Harris2021} and the square honeycomb \cite{Cote2004,Harris2017}. Between them, these architectures span a range of strut and plate orientations relative to the plane of build, which is an important parameter in AM processing. Specimens with each cellular geometry and process parameter combination were subjected to large strain compression (up to densification), quasi-statically and dynamically, with impact speeds up to 100 m/s, to assess the coupling between cellular architecture, process parameters, material defects and energy absorption performance.

The paper is structured as follows. In section 2, details are provided on the cellular specimen geometries and LPBF processing parameters. In section 3, material characterisation tests are carried out on solid dogbone specimens fabricated using identical LPBF processing parameters, to provide a base-line measure of the response of 316L stainless steel to these process variations. This includes microstructural characterisation, Vickers hardness testing and tensile testing. In section 4, the geometry and microstructure of the as-manufactured cellular specimens are characterised. The specimen densities are measured, and the feature resolution evaluated using X-ray CT scans and scanning electron microscopy (SEM). The quasi-static compression of the cellular specimens is presented in section 5. Performance metrics such as peak compressive strength, energy absorption up to densification and absorption efficiency are evaluated, and compared in the form of material property charts. Finally, in section 6, the dynamic compressive response of the cellular materials is described. A comparison is made between the quasi-static and dynamic load cases.

\section{Specimen Geometries and Processing Parameters}
\label{Sec:Specimen Geometries and Processing Parameters}

\subsection{Cellular Specimens}
\label{SubSec:Cellular Specimens}\vspace{2pt}

\begin{figure}[H]
	\centering
	\includegraphics[scale=0.4]{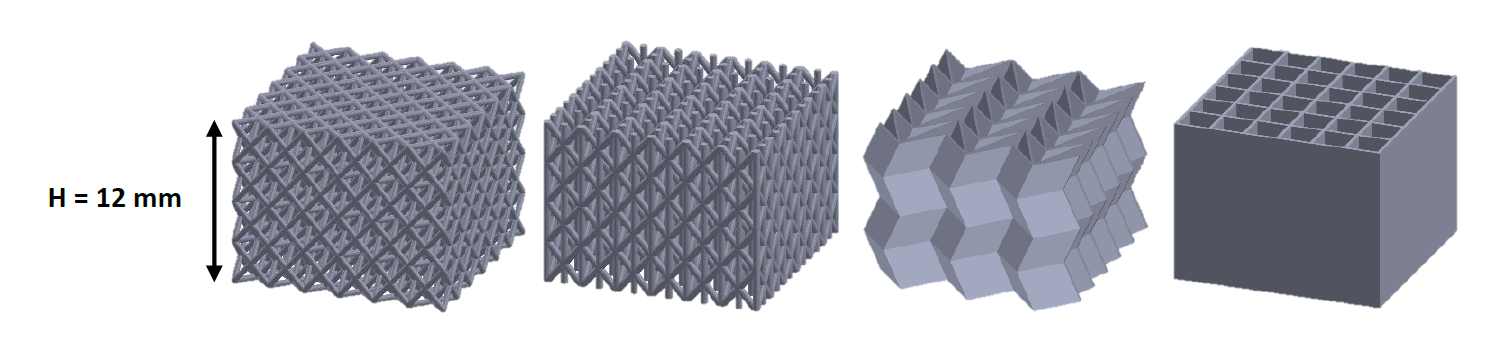}
	\caption{CAD models of the cellular geometries. From left to right: Octet Lattice, LW-SHC, Origami and SHC.}
	\label{Fig:CAD_Wv3}
\end{figure}

The four cellular material geometries considered in this investigation are shown in Fig. \ref{Fig:CAD_Wv3}. By spanning a range of cell wall types, connectivities and orientations, these provide contrast in terms of additive manufacturability and mechanical properties.

\begin{figure}[]
	\begin{subfigure}[h]{0.49\textwidth}
		\centering
		\includegraphics[scale=0.30]{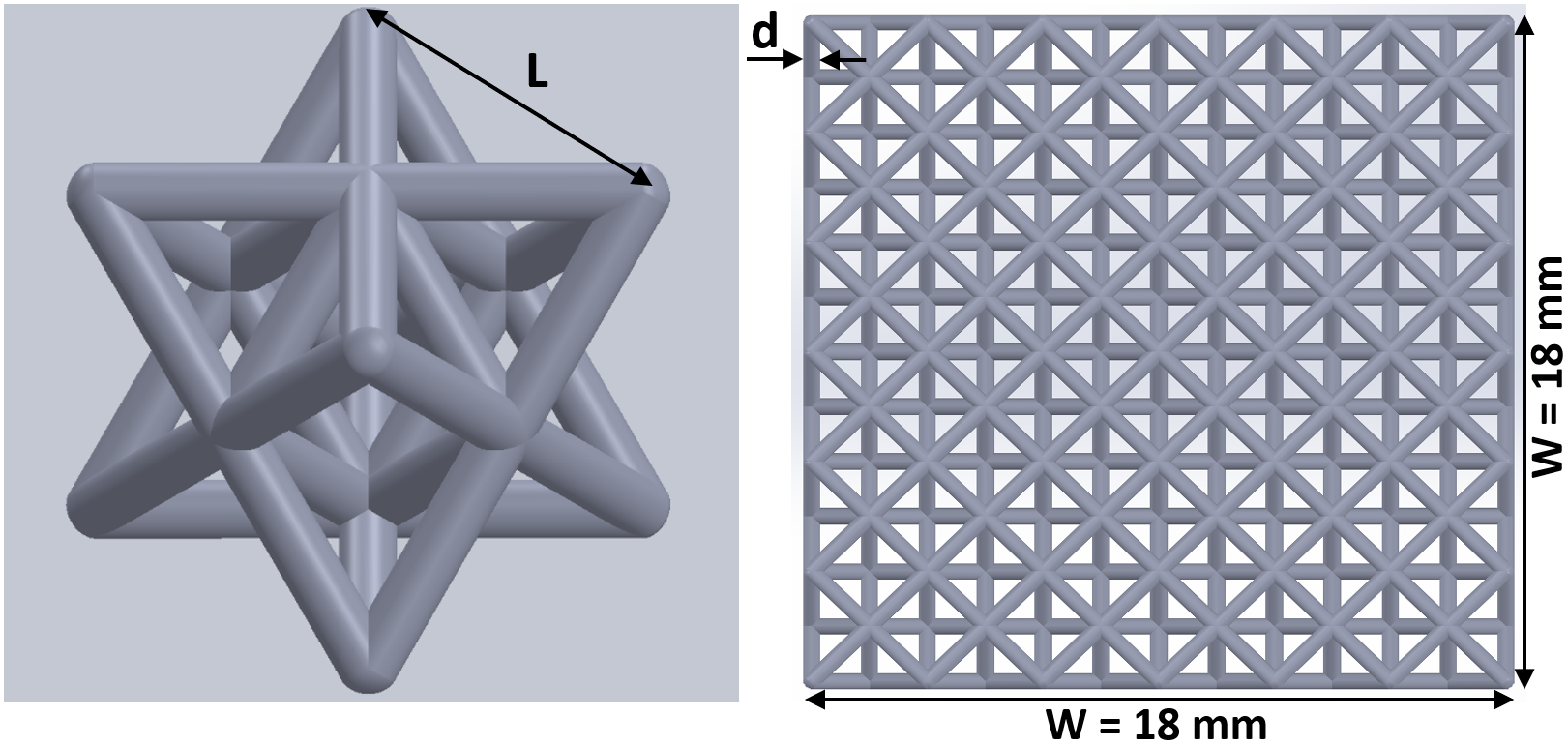}
		\caption{}
		\label{Fig:Octet_Combi_UnitCell}
	\end{subfigure}
	\begin{subfigure}[h]{0.49\textwidth}
		\centering
		\includegraphics[scale=0.30]{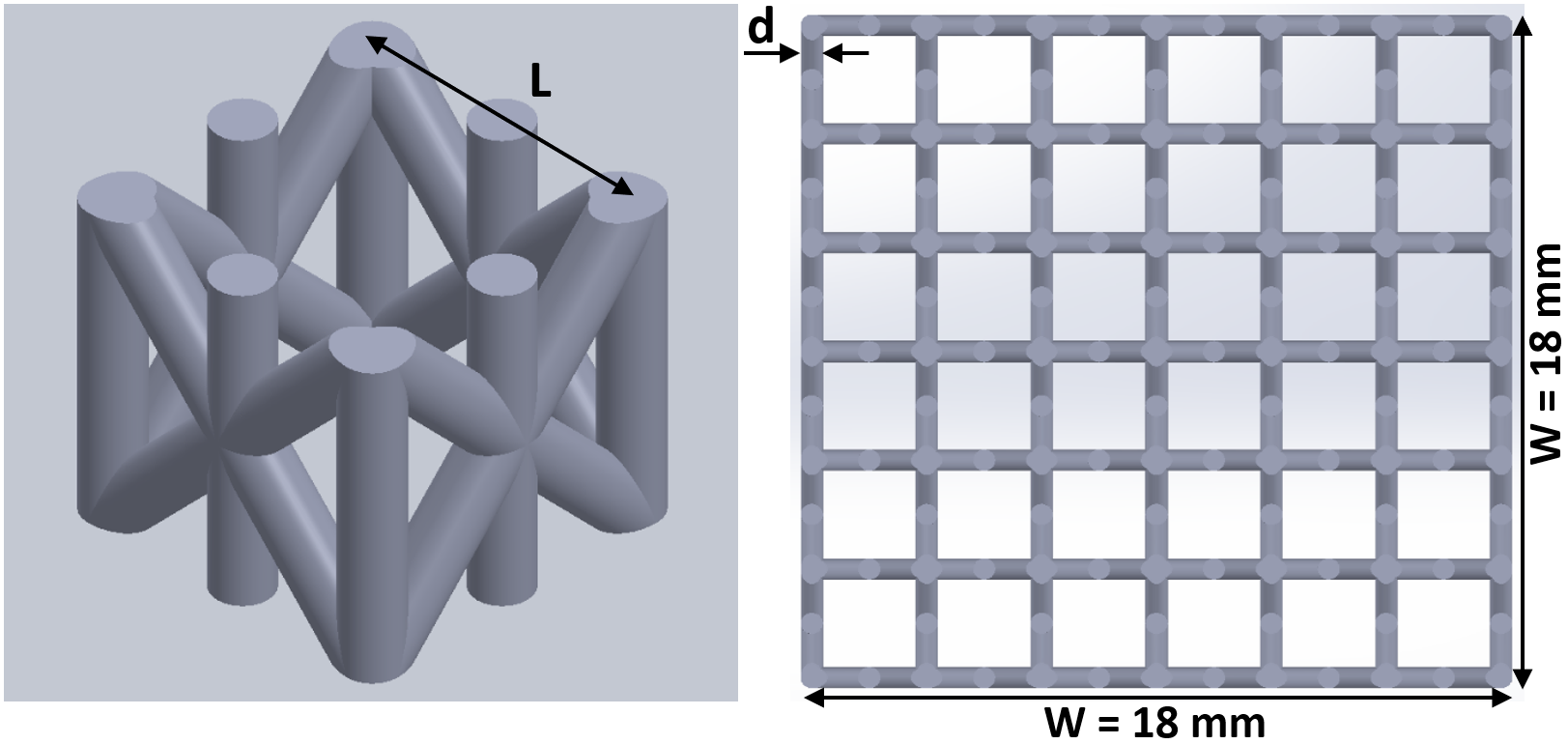}
		\caption{}
		\label{Fig:LWSHC_Combi_UnitCell}
	\end{subfigure}
	
	\begin{subfigure}[h]{0.49\textwidth}
		\centering
		\includegraphics[scale=0.30]{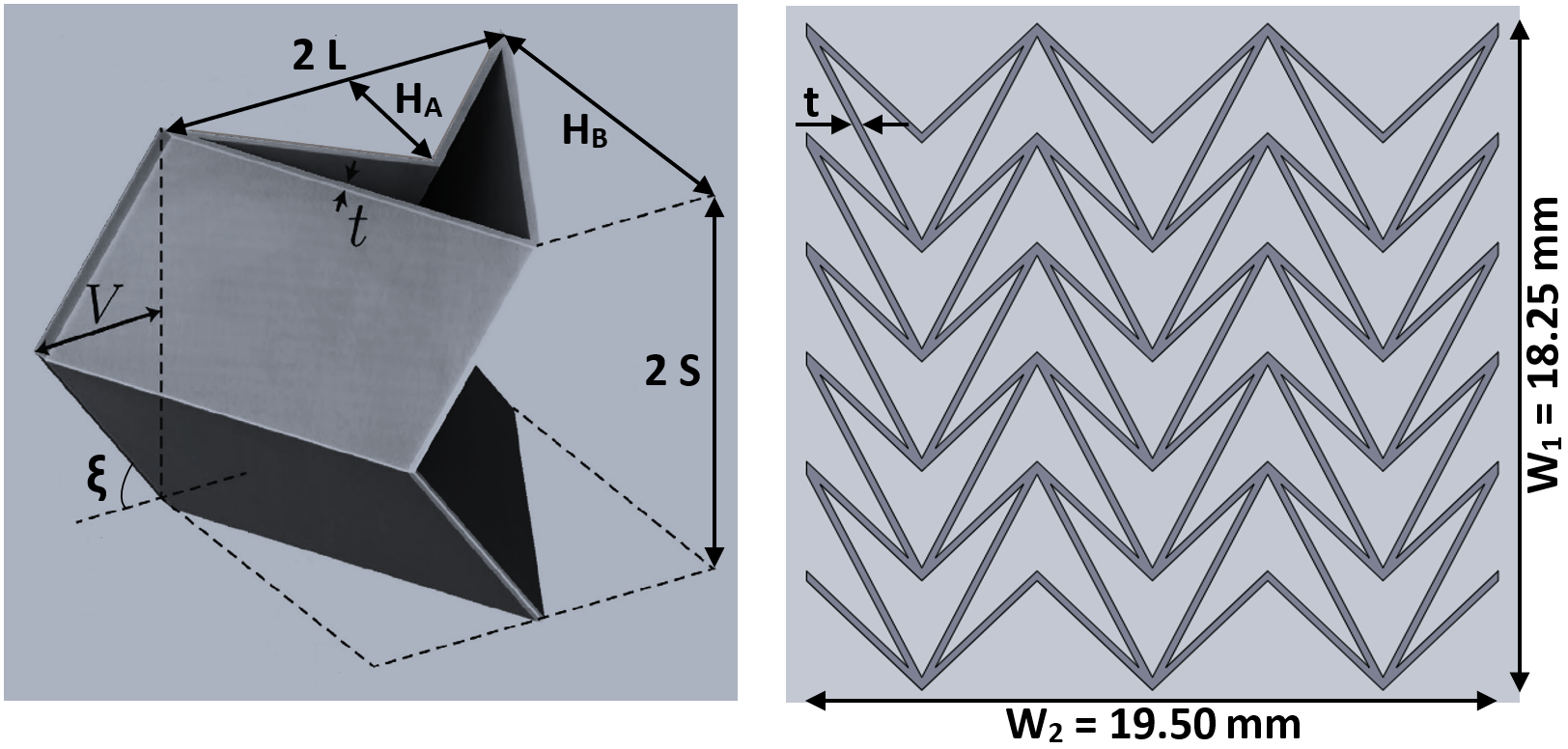}
		\caption{}
		\label{Fig:Origami_Combi_UnitCell}
	\end{subfigure}
	\begin{subfigure}[h]{0.49\textwidth}
		\centering
		\includegraphics[scale=0.30]{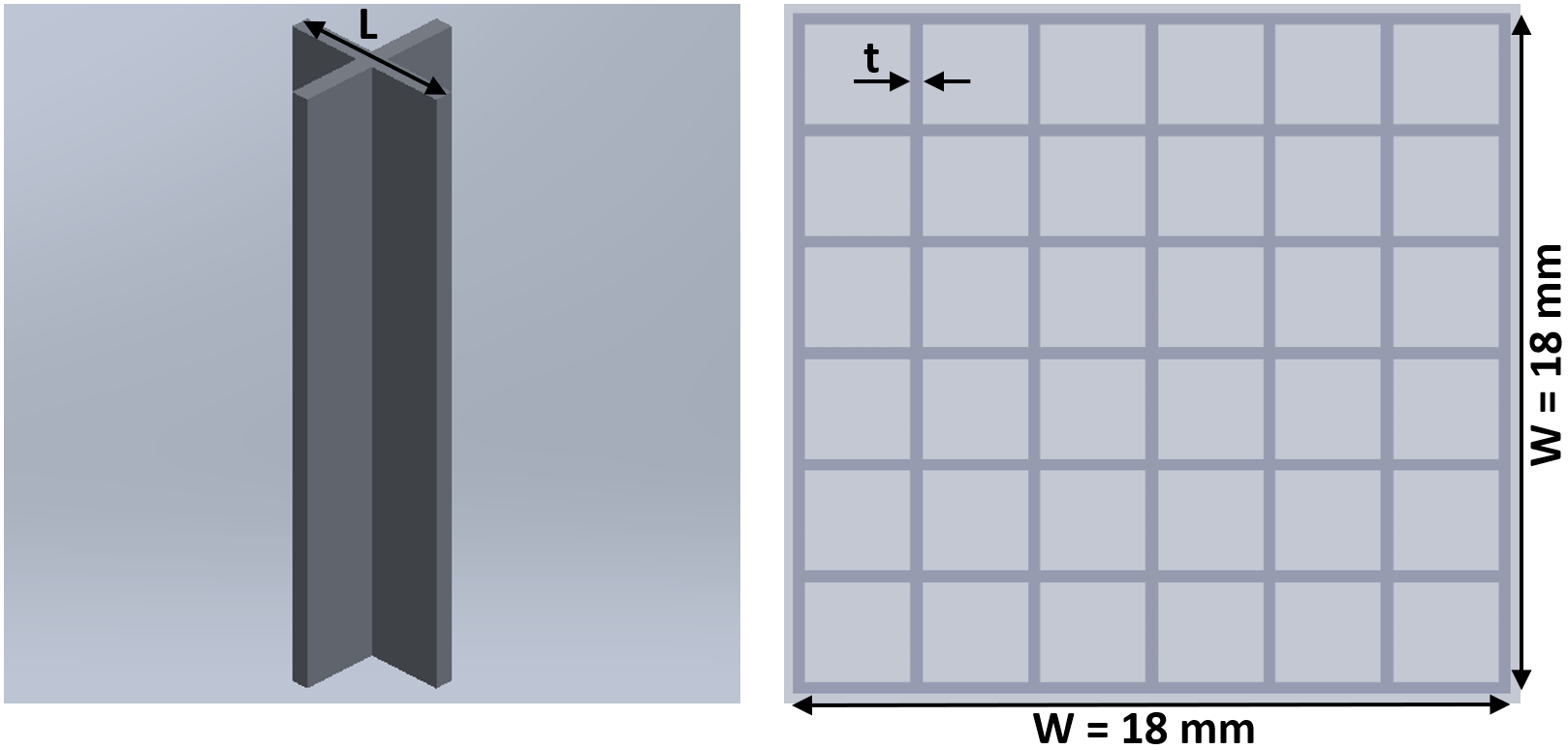}
		\caption{}
		\label{Fig:SHC_Combi_UnitCell}
	\end{subfigure}
	\caption{Unit cells (on the left) and plan views (on the right) of the cellular geometries: (a) octet truss lattice, (b) lattice-walled square honeycomb, (c) origami-inspired stacked Miura-ori and (d) square honeycomb. Note that the plan view is the cross-section parallel to the build plane. $L$ is the length of the unit cell, $d$ is the strut diameter, and $t$ is the wall thickness. $H_A$, $H_B$, $\xi$, $V$ and $S$ are additional parameters required to define the stacked Miura-ori fold pattern. Dimension $W$ is the side length of specimen.}
	\label{Fig:UnitCells}
\end{figure}

There are two lattice-based structures: the octet truss, and the lattice-walled square honeycomb (LW-SHC). The unit cells are shown in Fig. \ref{Fig:UnitCells}(a)-(b). The octet truss consists of a periodic array of bars with a nodal connectivity of 12, the lower limit for a three dimensional stretching dominated lattice with similarly situated nodes \cite{Deshpande2001a}. The LW-SHC was proposed by Harris et al. \cite{Harris2017a} as a hybrid structure: it can be considered a square honeycomb with the solid facets replaced with a triangulated lattice. It has a nodal connectivity of 6 where struts meet in the centres of the `honeycomb' walls, and 10 at the intersections between walls.

The remaining two cellular geometries are facet-based, i.e.\ they consist of a periodic array of plates, rather than bars: the origami-inspired stacked Miura-ori and the square honeycomb (SHC). The stacked Miura-ori is of interest for its property tunability: the use of origami fold patterns lends it a very wide cell geometry design space, and makes it suitable for graded and curved cellular structures \cite{Harris2020,Harris2021}. The Miura-ori fold pattern is flat-foldable, which makes the stacked Miura-ori a bending dominated structure for out-of-plane loading (for further details refer to \cite{Schenk2013,Schenk2014}). The SHC is a stretching dominated structure, and demonstrates a high specific stiffness and strength, when compressed out-of-plane \cite{Cote2004}. The unit cells are shown in Fig. \ref{Fig:UnitCells}(c)-(d).

Fig. \ref{Fig:CAD_Wv3} shows the sizes of the cellular specimens as manufactured. The height of all specimens is $H$ = 12 mm. The dimension $H$ is oriented perpendicular to the build plane in all cases. The plan views of the specimens as manufactured are given in Fig. \ref{Fig:UnitCells}. Each cellular specimen is designed to fit within a bounding footprint of approximately 18 mm x 18 mm, which was set by the constraints of the mechanical testing apparatus (described subsequently). Each specimen has approximately 6 x 6 unit cells within this footprint. It is desirable to have as many cells as possible in each dimension so as to minimise edge effects (cells on the free surfaces have reduced connectivity compared to interior cells). However, there is a trade-off between cell size and relative density, as the LPBF process introduces a minimum strut diameter and facet thickness. The nominal cell sizes, wall thicknesses and other cell geometry parameters are given in Table \ref{Tab:Parameters_Geometries}. These dimensions match the CAD models used to build the AM part.

\begin{table}[H]
	\centering
	\caption{Nominal dimensions of the cellular specimens. The geometry parameters are defined in Fig. \ref{Fig:UnitCells}. Further details on the geometry of the origami-based stacked Miura-ori can be found in \cite{Harris2020,Harris2021}.}
	\label{Tab:Parameters_Geometries}
		\begin{tabular}{|c|c|c|c|c|c|c|c|c|}
			\hline
			Specimens & Cell Counts & \begin{tabular}[c]{@{}c@{}}$t$\\ {[}$mm${]}\end{tabular} & \begin{tabular}[c]{@{}c@{}}$d$\\ {[}$mm${]}\end{tabular} & \begin{tabular}[c]{@{}c@{}}$L$\\ {[}$mm$\end{tabular} & \begin{tabular}[c]{@{}c@{}}$H_A$\\ {[}$mm${]}\end{tabular} & \begin{tabular}[c]{@{}c@{}}$H_B$\\ {[}$mm${]}\end{tabular} & $\left(\frac{V}{S}\right)$ & \begin{tabular}[c]{@{}c@{}}$\xi=tan^{-1}\left(\frac{V}{S}\right)$\\ {[}$radians${]}\end{tabular} \\ \hline
			Octet Lattices & 6 x 6 x 4 & - & 0.45 & 3 & - & - & - & - \\ \hline
			LW-SHC & 6 x 6 x 4 & - & 0.58 & 3 & - & - & - & - \\ \hline
			Origami & 5 x 5 x 2 & 0.18 & - & 3 & 3 & 1.5 & 0.52 & 0.48 \\ \hline
			SHC & 6 x 6 x 1 & 0.32 & - & 3 & - & - & - & - \\ \hline
		\end{tabular}
\end{table}

The cell geometry parameters were chosen to give a nominal relative density $\bar{\rho}$ = 0.20 in all cases. Here, $\bar{\rho}$ is defined as the volume fraction of solid for an infinite array of cells. As described subsequently, a specimen with a finite number of cells can deviate from this infinite array volume fraction due to edge effects, i.e.\ truncation of cells at the free surfaces, and how this is accounted for. Here, for consistency, we use the infinite array value to specify the cell geometry parameters. The chosen value of $\bar{\rho}$ is determined by the minimum resolvable feature size, once the number of cells across the footprint area has been chosen. For reference, under `ideal' manufacturing conditions, this infinite cellular array would have a density \cite{Gibson1997}:
\begin{equation} \label{Eq:Rhoinfty}
\rho_{\infty} = \bar{\rho} \; \rho_s = 1600 \, \text{kg/m}^{3}
\end{equation}
\noindent where $\rho_s$ = 8000 kg/m$^3$ is the density of conventionally processed 316L stainless steel \cite{ASM316L}.

Ghouse et al. \cite{Ghouse2017b} demonstrated that, for LPBF processes, the resolution of lattice struts is affected by the minimum angle with respect to the build plane. Struts oriented at a low angle can experience manufacturing problems. This affects the octet lattice, where some struts are nominally oriented parallel to the build plane. Therefore two variants of the octet lattice are considered: the original geometry (denoted the 0$^{\circ}$ case), and a version with the horizontal struts rotated by 15$^{\circ}$ relative to the build plane.

\subsection{LPBF Process Parameters}
\label{SubSec:LPBF Manufacturing Parameters}\vspace{2pt}

The specimens were manufactured using a Renishaw AM250 metal powder bed fusion additive manufacturing system, equipped with a Gaussian beam fibre laser. Full details of this AM system can be found in Ghouse et al. \cite{Ghouse2017b}, and are reproduced here for reference.  The laser has a maximum power of 200 W and a wavelength of $\lambda$ = 1070 nm, and it produces a spot size of 70 $\mu$m. The AM250 is a modulated continuous wave system where the laser fires at a point for a given exposure time before moving on to the next point. Once the exposure time ends, it turns off and moves to the next point by a set point distance. This method is repeated along each scan vector. The manufacturing process was initially performed in a vacuum (-960 mbar) and afterwards back-filled with 99.995$\%$ pure Argon to 10 mbar with 0.1$\%$ Oxygen. The 316L stainless steel powder consisted of spherical particles of size range 10-45 $\mu$m (D$_{50}$: 27 $\mu$m), and was supplied by LPW Technology. The specimens were removed from the substrate using electro discharge machining (EDM) and were cleaned ultrasonically in ethanol.

All specimens were built using a contour laser scanning strategy, as detailed in Ghouse et al. \cite{Ghouse2017b}. A fixed distance of 50 $\mu$m between each exposure point was selected, and the contour spacing was 35 $\mu$m. Three process parameter sets were considered that alter the nature of the heat input. Three different laser power values were chosen, and in each case the exposure time was adjusted to keep a fixed total heat input (i.e.\ a fixed specific enthalpy, as defined in \cite{Ghouse2017b}). The combinations of laser power and exposure time for process parameter sets 1, 2 and 3 are given in Table \ref{Tab:SLMParameters}.

\begin{table}[H]
	\centering
	\caption{LPBF process parameter sets.}
	\label{Tab:SLMParameters}
	\begin{tabular} {|c|c|c|}
		\hline
		\begin{tabular}[c]{@{}l@{}}Parameter Set\end{tabular} & \begin{tabular}[c]{@{}l@{}}Laser Power [$W$]\end{tabular} & \begin{tabular}[c]{@{}l@{}}Exposure Time [$\mu$s]\end{tabular} \\ \hline
		1 & 50 & 190 \\
		2 & 125 & 90 \\
		3 & 200 & 40 \\
		\hline
	\end{tabular}
\end{table}

\section{Constituent Material Response to the AM Process Variants}
\label{Sec:Influence AM on Dogbones}

To provide a baseline for understanding the impact of the process variants, independent of any cellular architecture, dogbone tensile test specimens are manufactured from 316L stainless steel using the three LPBF process variants (parameter sets 1, 2, and 3). The dogbones are used to assess differences in microstructure and mechanical properties between the process variants.

\begin{figure}[htp]
	\centering
	\includegraphics[scale=0.4]{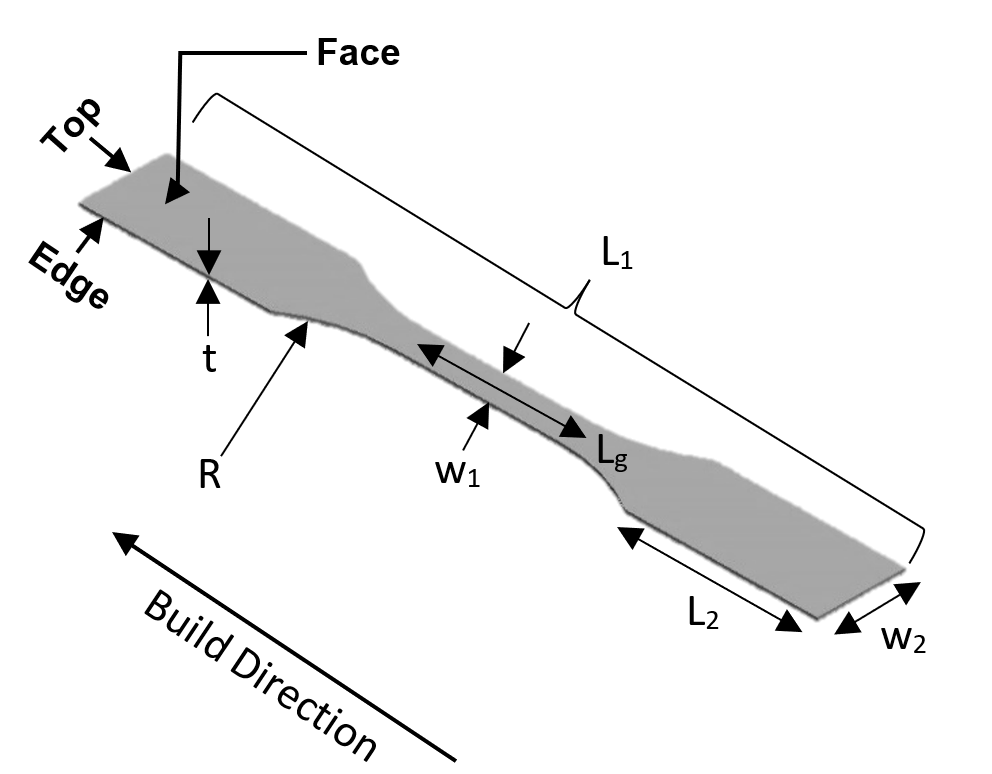}
	\caption{Nominal geometry of the solid dogbones with the build direction for LPBF manufacturing indicated. The top, edge and face views highlighted were considered for microstructure inspection. }
	\label{Fig:Dogbone_v2}
\end{figure}

The geometry of the dogbone specimens are shown in Fig. \ref{Fig:Dogbone_v2}. The nominal dimensions are: total length $L_1 = 65$ mm; grip section length $L_2 = 16.25$ mm; gauge section length, $L_g = 12.5$ mm; gauge section width, $w_1 = 3$ mm; grip section width $w_2 = 7$ mm; thickness, $t = 1$ mm; radius $R = 22.5$ mm. The stress uniformity within the gauge length for this dogbone geometry was verified through finite element analysis. The dogbones were manufactured with the gauge section oriented perpendicular to the build plane, as this configuration showed the best mechanical performance and provided microstructure characteristics closer to that of the cellular materials \cite{Harris2017a}.

\subsection{Microstructural Characterisation}
\label{SubSec:Microstructural Characterisation of Solid Dogbones}\vspace{2pt}

For the microstructural inspection of the dogbones, specimens were cut from the grip sections (prior to tensile testing), mounted, polished, and etched with Glycerygia (15 ml glycerol, 10 ml HCl, 5 ml HNO$_3$) for 60 - 75 seconds. Images were taken using dark field optical microscopy at different levels of magnification.  Surfaces in three orientations were inspected, denoted `top', `edge' and `face' views, as shown in Fig. \ref{Fig:Dogbone_v2}.

\begin{figure}[htp]
	\centering
	\includegraphics[scale=0.8]{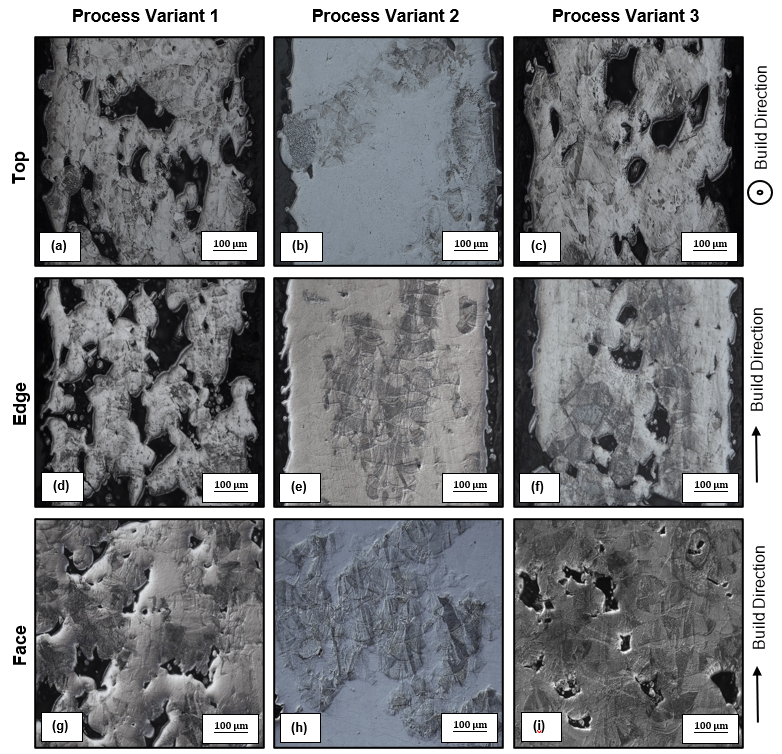}
	\caption{Microstructure of solid dogbones observed using optical microscopy at a magnification of 10x: (a-c) top view, (d-f) edge view, (g-i) face view.}
	\label{Fig:DogboneMicrostructure1_v2}
\end{figure}

\begin{figure}[H]
	\centering
	\includegraphics[scale=0.8]{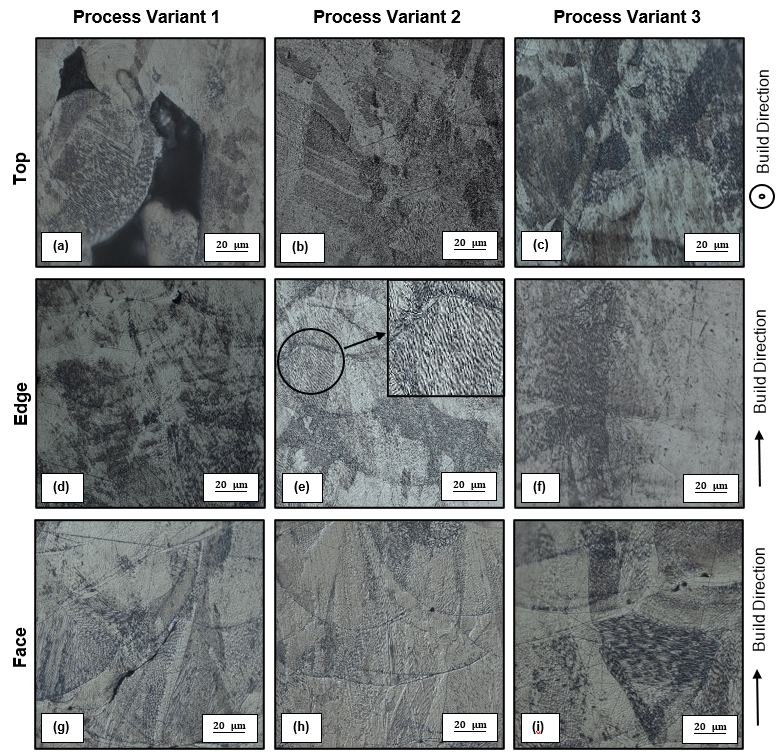}
	\caption{Microstructure of solid dogbones observed using optical microscopy at a magnification of 50x: (a-c) top view, (d-f) edge view, (g-i) face view.}
	\label{Fig:DogboneMicrostructure2_v2}
\end{figure}

Micrographs at 10x magnification are shown in Fig. \ref{Fig:DogboneMicrostructure1_v2}. Significant porosity is present in all three views of the dogbones manufactured with the lowest and highest laser powers (50 W and 200 W, process variants 1 and 3 respectively); the porosity with process variant 1 appears slightly higher. In comparison, almost no porosity is apparent in the dogbones fabricated with process parameter set 2 (125W, the intermediate laser power). Micrographs at 50x magnification are shown in Fig. \ref{Fig:DogboneMicrostructure2_v2}, revealing the grain structure. The curved features, particularly visible in the face view, Fig. \ref{Fig:DogboneMicrostructure2_v2}(g-i), indicate the melt pool boundaries \cite{Yusuf2017}. The size of these features is comparable with the 70 $\mu$m laser spot size. Fine scale sub-grain solidification structures approximately 0.1-2 $\mu$m in size (see the magnified image in Fig. \ref{Fig:DogboneMicrostructure2_v2}(e)) result from the rapid solidification rates in the locally melted areas of material \cite{Yusuf2017}. The grain size, the shape of the melt pool boundaries and the fine sub-grain solidification structures appear to be insensitive to the process variant. The key microstructural impact of the process variants is therefore at the level of the porosity.

There have been a number of investigations into the origins of porosity in LPBF processes \cite{Sames2016a, Sames2014a, King2014, Tang2017, Yusuf2017}. Pores of a spherical shape can indicate nucleation of gas bubbles. However, the large scale porosity in Fig. \ref{Fig:DogboneMicrostructure1_v2} for process parameter sets 1 and 3 appears more irregular in shape. Irregular pores can be related to incomplete melting or fusion of powder particles, as a result of insufficient energy being supplied (through the choice of laser power, or insufficient overlap of successive melt pools). This is a reasonable explanation for the porosity observed in parameter set 1, where the energy is applied at the lowest power over the longest exposure time. An incompletely fused powder particle is apparent in Fig. \ref{Fig:DogboneMicrostructure2_v2}(a). Irregularly shaped porosity can also result from excessive energy, which results in a spatter ejection \cite{Yusuf2017}. This can account for the porosity in parameter set 3, the highest power level. Process parameter set 2 achieves a low porosity by avoiding either incomplete fusion or spatter ejection.

\subsection{Porosity Measurement}
\label{SubSec:Porosity of the Solid Dogbones}\vspace{2pt}

Next, a quantitative analysis of this porosity is carried out. ImageJ image processing software \cite{Schneider2012b} was used to measure the area fraction of pores visible in the micrographs. The porosity estimate from the ImageJ analysis ($p_J$) is given by
\begin{equation}\label{Eq:PorosityImageJ}
p_J = \frac{A_p}{A}
\end{equation}
\noindent where $A_p$ is the area of pores in a given image, and A is the total area of the image. In order to verify these values, a second estimate of the porosity ($p_D$) was obtained from a measurement of the density of the dogbone specimens. The mass of the dogbones was obtained using an Oertling R20 balance (1 division = 1 mg). Measurements were taken of the length ($L_1$ and $L_2$), gauge length ($L_g$), and width ($w_1$ and $w_2$), as well as of the thickness ($t$), of the as-manufactured dogbones. By assuming a fillet radius of 22.5 mm (as in the CAD model), these measurements are used to estimate the as-manufactured volumes of the dogbones. 
An estimate of the material density $\rho$ is found by dividing the measured mass by this approximate as-manufactured dogbone volume. The porosity is then given by
\begin{equation}\label{Eq:Porosity}
p_D = 1-\frac{\rho}{\rho_0}
\end{equation}
\noindent where $\rho_0$ = 8000 kg/m$^3$ corresponds to fully dense conventionally processed 316L stainless steel \cite{ASM316L}.

The porosity and density estimates for the dogbones manufactured using each of the process variants are given in Table \ref{Tab:Densities+Porosity}. The porosity values obtained using the two techniques reflect the qualitative trends from the micrographs. Process parameter sets 1 and 3 show significantly higher porosity levels compared to process variant 2.

\begin{table}[H]
	\centering
	\caption{The estimated material density, the porosity obtained using ImageJ image processing ($p_J$) and the porosity obtained using the density measurements ($p_D$).}
	\label{Tab:Densities+Porosity}
		\begin{tabular}{|c|c|c|c|}
			\hline
			Parameter Set & Material Density, $\rho$ [$kg/m^3$] & $p_J$ & $p_D$ \\ \hline
			1               & 6405    & 0.29                               & 0.20 \\ \hline
			2               & 7245    & 0.02                               & 0.09 \\ \hline
			3               & 6693    & 0.17                               & 0.16 \\ \hline
		\end{tabular}
\end{table}

\subsection{Vickers Hardness Tests}
\label{SubSec:Vickers Hardness Tests of Solid Dogbones}\vspace{2pt}

\begin{figure}[htp]
	\centering
	\includegraphics[scale=0.6]{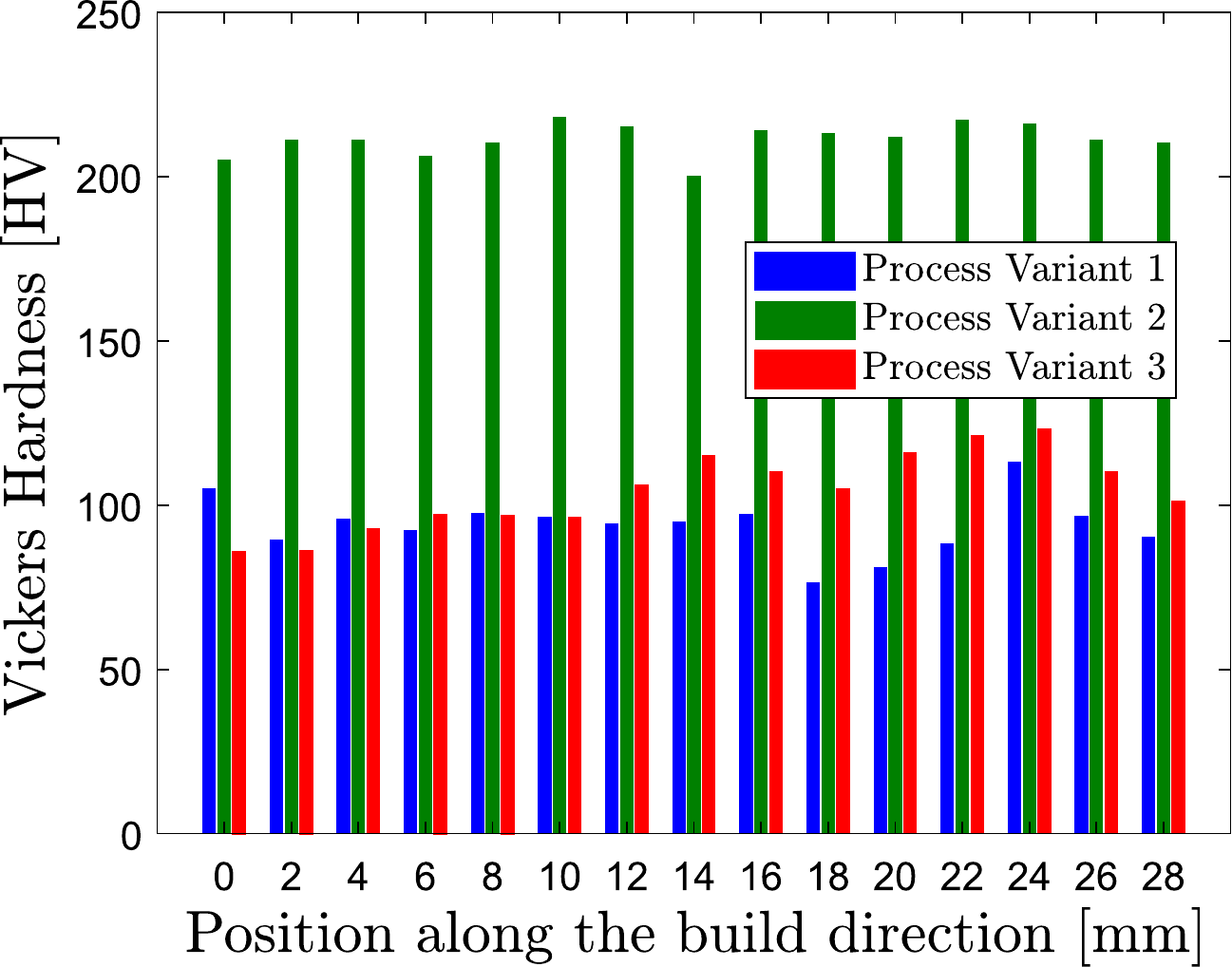}
	\caption{Vickers hardness test results for the solid dogbones manufactured with the three different process parameter sets. Hardness measurements were taken in 15 locations along the total length of the dogbones' face, in a line parallel to the build direction. The distance between each measurement was 2 mm.}
	\label{Fig:VickersHardnessHV}
\end{figure}

Vickers hardness tests were conducted on the dogbone specimens, to determine any variations in yield strength in the build direction, i.e.\ along the length of the dogbone. A Vickers-Armstrongs hardness testing machine was used with a 136$^\circ$ diamond pyramid indenter and a load of 30 kg. The measurements were taken on the face of the dogbones at 2 mm intervals along a line parallel to the build direction (see Fig. \ref{Fig:Dogbone_v2}). The results are shown in Fig. \ref{Fig:VickersHardnessHV}.

A uniform hardness along the length of the dogbone is observed for all process variants. Small variations may be attributable to the location of pores relative to the indent. Note that the average hardness value of 211 HV measured for the process parameter set 2 specimen is approximately 36\% higher than the AISI data sheet value of 155 HV \cite{ASM316L}. Yusuf et al. \cite{Yusuf2017} also obtained microhardness values higher than 200 HV for LPBF processed 316L stainless steel. This can be attributed to the fine-scale microstructure described in Section \ref{SubSec:Microstructural Characterisation of Solid Dogbones}, which results in material with greater hardness and tensile strength compared to wrought values \cite{Yasa2011a, Martinez2012, Yusuf2017}. However, the hardness values for the dogbones produced using process parameter sets 1 and 3 are around half those of process variant 2. This is a result of the higher porosity levels.

\subsection{Tensile Tests}
\label{SubSec:Tensile Tests of Solid Dogbones}\vspace{2pt}

Tensile tests were carried out on the dogbones using an Instron screw-driven test machine (150 kN load cell) at a nominal strain rate of 10$^{-3}$ s$^{-1}$. The deformation of the gauge section was measured using Digital Image Correlation (DIC). A clip gauge was used to provide high resolution data at small strains, and a laser extensometer verified the DIC measurements at larger strains. The stress in the gauge section was calculated from the test machine load cell measurements.

\begin{figure}[H]
	\begin{subfigure}[!h]{0.49\textwidth}
		\centering
		\includegraphics[scale=0.6]{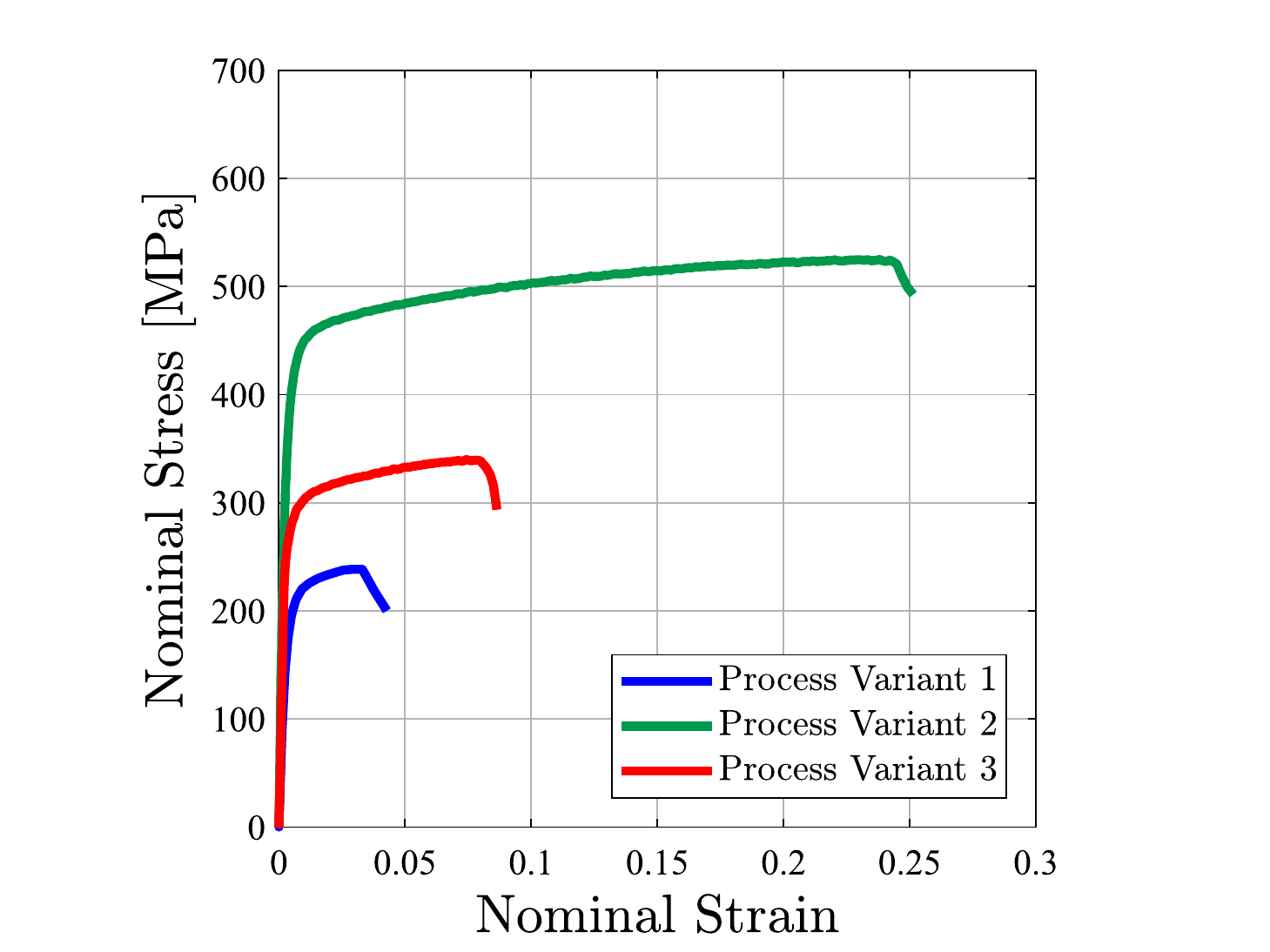}
		\caption{}
		\label{Fig:DIC_NominalStressNominalStrain}
	\end{subfigure}
	\begin{subfigure}[!h]{0.49\textwidth}
		\centering
		\includegraphics[scale=0.6]{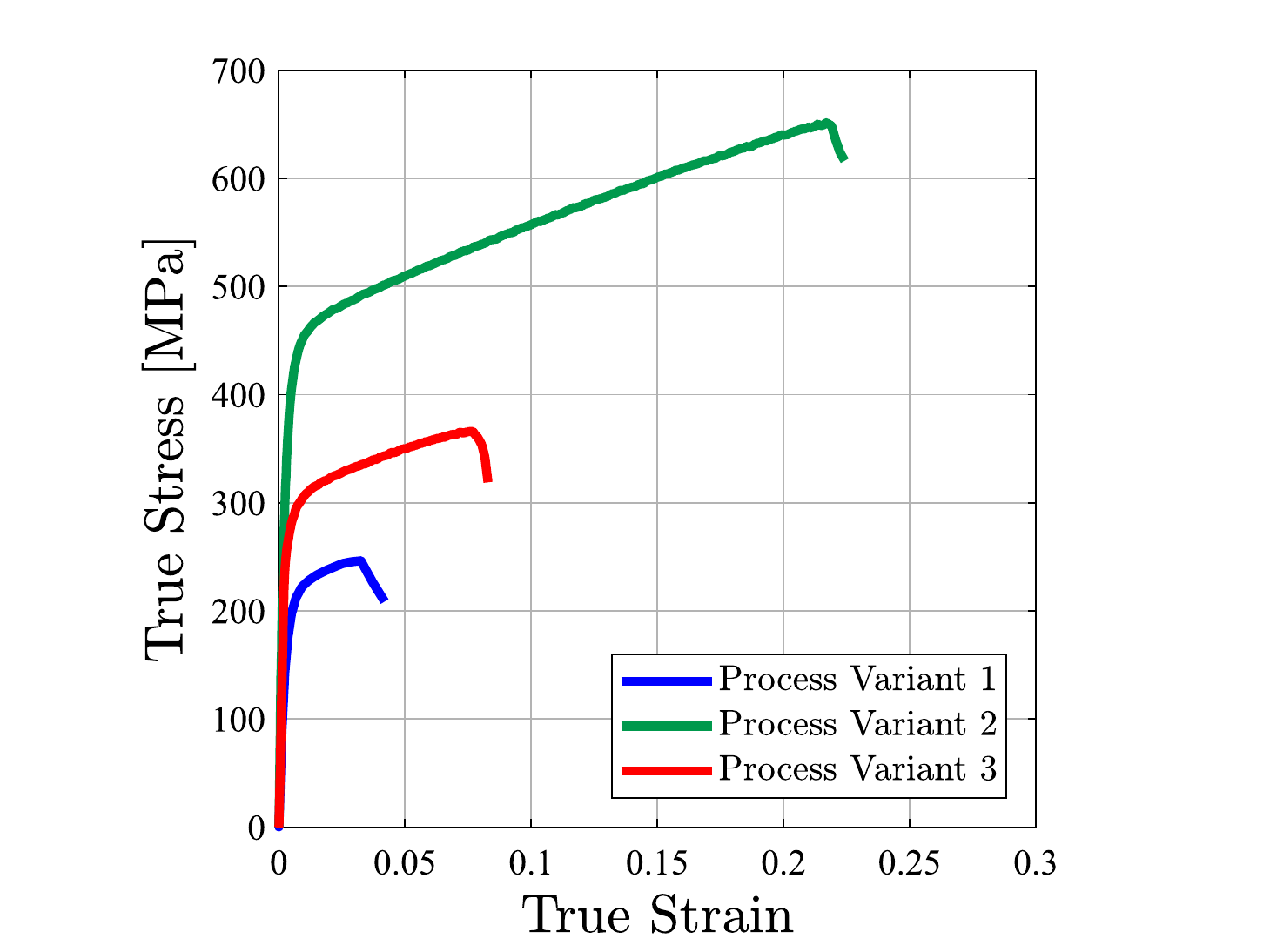}
		\caption{}
		\label{Fig:DIC_TrueStressTrueStrain}
	\end{subfigure}
    	
    \caption{(a) Nominal stress versus nominal strain and (b) true stress versus true strain measurements for the LPBF 316L stainless steel dogbone specimens manufactured using the three process parameter sets/variants.}
    \label{Fig:DIC_StressStrain}
\end{figure}

The nominal tensile stress-strain curves for dogbones manufactured with the three process variants are plotted in Fig. \ref{Fig:DIC_StressStrain}(a). The nominal stress was obtained from the measured initial cross-sectional area of the gauge section. The nominal strain was obtained from the DIC results, tracking the separation of two points within the gauge section. The DIC also confirmed uniform straining within the gauge section prior to the ultimate tensile strength. In all cases, necking and fracture occurred within the gauge section. Fig. \ref{Fig:DIC_StressStrain}(b) shows the true stress-strain curves for the three dogbones, calculated by assuming conservation of volume during plastic deformation. Given the measured porosity, this assumption was checked using the DIC data. For process parameter sets 2 and 3, measurement of the transverse and longitudinal strains confirmed a plastic Poisson's ratio $\nu_p \approx 0.5$. The specimen produced using process variant 1 failed at a much smaller tensile strain, and a value for the plastic Poisson's ratio was not obtained. However, due to the small strain level, there is negligible difference between the nominal and true stress for this specimen. The true stress-strain curves show a similar tangent modulus for all three process variants, with value $E_t \approx 900$ MPa.

In order to measure the Young's modulus, the samples were unloaded and reloaded in the elastic region, and at several points along the stress-strain curve, with a clip gauge used to provide high resolution strain data. The measured Young's modulus values are given in Table \ref{Tab:Modulus_Strength}. The values are substantially lower than the bulk annealed value for 316L stainless steel, $E \approx$ 193-200 GPa \cite{ASM316L}. However, comparable values for LPBF 316L stainless steel have been reported in other investigations, with the low values attributed to porosity and small cracks found between the melt layers \cite{Correa2015, Geenen2017}. Comparing the three process parameter sets, increased porosity (Table \ref{Tab:Densities+Porosity}) corresponds to a lower Young's modulus.

The 0.2$\%$ offset yield strengths for the three dogbone specimens are given in Table \ref{Tab:Modulus_Strength}. For reference, the AISI annealed value for 316L stainless steel is $\sigma_y = $ 290 MPa \cite{ASM316L}. As observed with the Vickers hardness measurements, process variant 2 has the highest yield strength, followed by variants 3 and 1.  The yield strength for process parameter set 2 is 26\% higher than the AISI annealed value. As noted above, this can be attributed to the fine-scale microstructure of the LPBF material. Similar results were obtained for LPBF-processed material by Shen et al. \cite{Shen2010}. The reduced yield strengths and ductilities of the specimens produced using process parameter sets 1 and 3 can be attributed to their higher porosity level. Comparing the yield strengths with the measured porosity values, the following relationship fits the data well:
\begin{equation}\label{Eq:PowerLawStrength}
\sigma_y=\sigma_{y0}(1 - p_J)^{n}
\end{equation}
with the fitting constants taking the values $\sigma_{y0} = 375$ MPa and $n = 2$.

\begin{table}[H]
	\centering
	\caption{Young's modulus and 0.2$\%$ offset yield strength results from the tests carried out on the dogbones manufactured with the three process parameter sets.}
	\label{Tab:Modulus_Strength}
		\begin{tabular}{|c|c|c|}
			\hline
			Parameter Set & Young's Modulus, $E$ {[}GPa{]} & 0.2\% Offset Yield Strength, $\sigma_y$ {[}MPa{]} \\ \hline
			1 & 92 & 190 \\ \hline
			2 & 168 & 365 \\ \hline
			3 & 100 & 250 \\ \hline
		\end{tabular}
\end{table}

\section{As-manufactured Cellular Specimens}
\label{Sec:Cellular Specimens: geometry, microstructure and defects.}

\begin{figure}[H]
	\centering
	\includegraphics[scale=0.7]{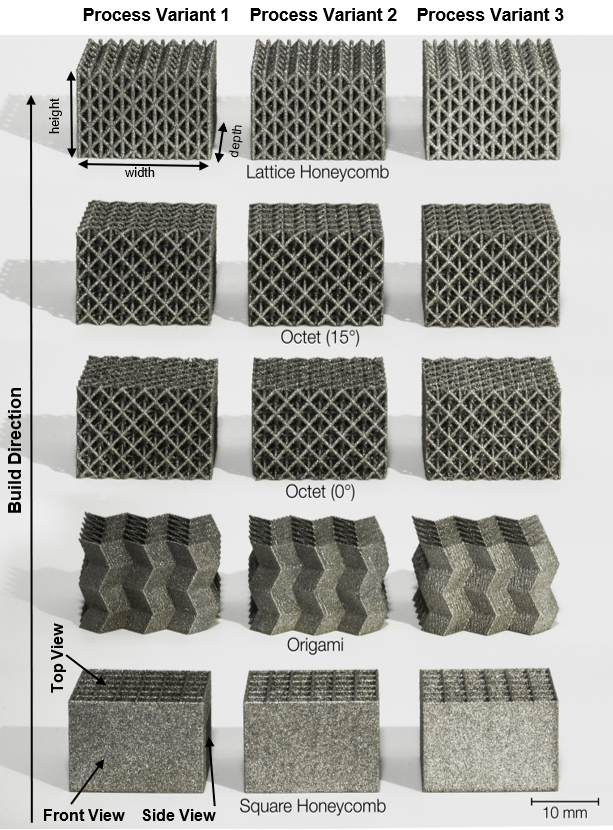}
	\caption{Photographs of the five cellular geometries as-manufactured with three LPBF process parameter sets.}
	\label{Fig:FifteenSpecimens_v2}
\end{figure}

In this section, the feature resolution, density and microstructure of the as-manufactured LPBF cellular specimens are described. Photographs of the as-manufactured specimens are shown in Fig. \ref{Fig:FifteenSpecimens_v2}. There are fifteen specimen types in total: five cellular geometries (including the two versions of the octet truss), each produced using the three LPBF process variants.

\subsection{Density Measurements}
\label{SubSec:Macro-Level Analysis}\vspace{2pt}

\begin{figure}[htp]
	\centering
	\includegraphics[scale=0.7]{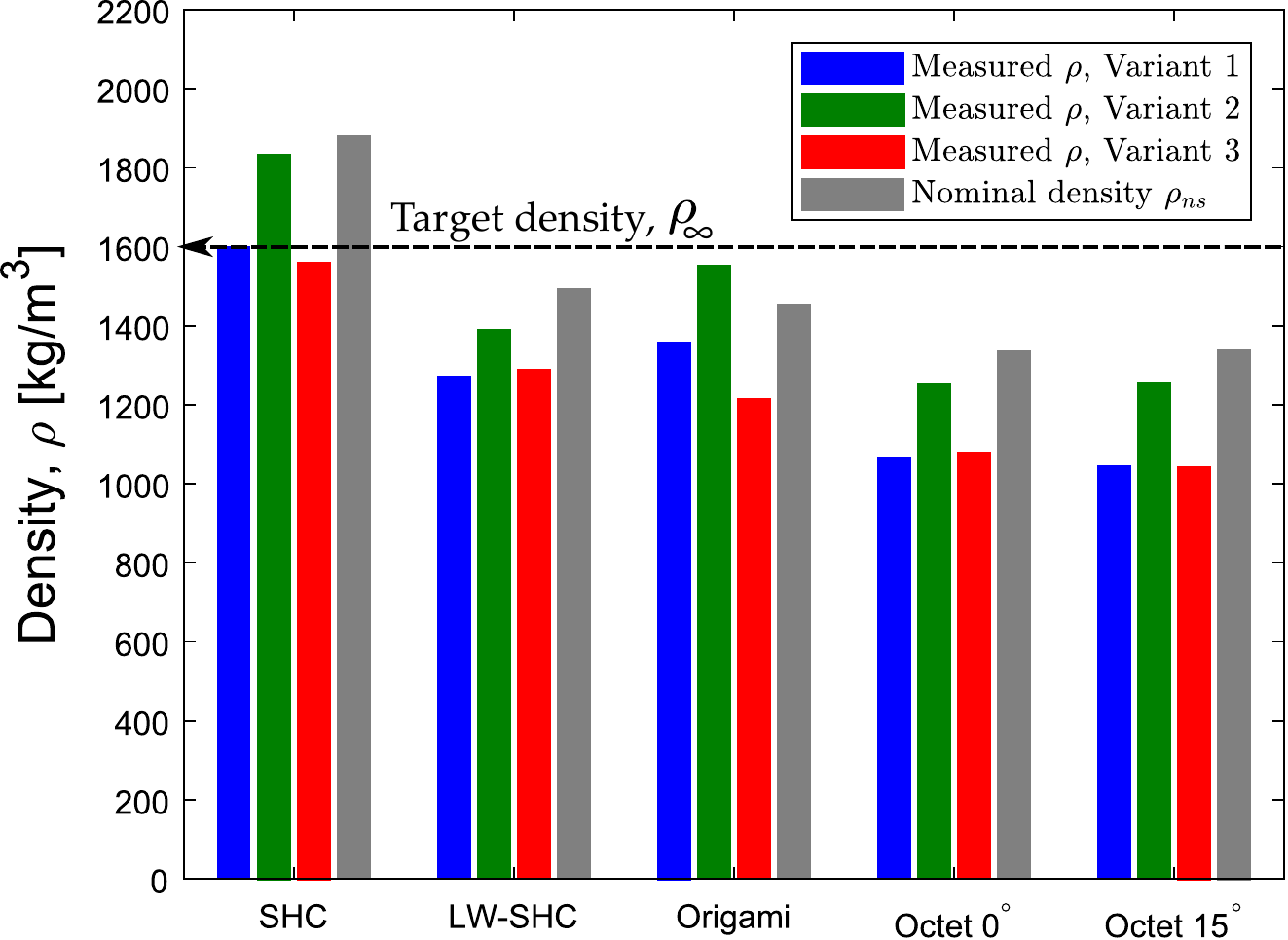}
	\caption{Densities ($\rho_{\infty}$, $\rho_{ns}$, and measured $\rho$) of the five cellular geometries as manufactured with the three LPBF process variants.}
	\label{Fig:ThreeDensityCellular22Jan20}
\end{figure}

The measured density ($\rho$) of the as-manufactured cellular specimens was obtained by dividing the specimen mass by the volume of their measured bounding box (i.e. the footprint area multiplied by the height, approximately 18x18x12 mm$^3$ (19.50x18.25x12.00 mm$^3$ for the origami), shown in Figs. \ref{Fig:CAD_Wv3} and \ref{Fig:UnitCells}. The results are presented in Fig. \ref{Fig:ThreeDensityCellular22Jan20}.

For comparison, the nominal specimen density values $\rho_{ns}$ are also given, calculated as follows. Using the CAD models of the cellular specimens, with the nominal dimensions, the volume fraction of solid is calculated for the same bounding box used for the experimentally measured densities. Multiplying this nominal solid volume fraction by $\rho_s$ = 8000 kg/m$^{3}$, for fully dense 316L stainless steel \cite{ASM316L}, gives $\rho_{ns}$. Note that these `specimen nominal densities' deviate from the `infinite array' target density value $\rho_{\infty}$ used in the specimen design, as described in Section \ref{SubSec:Cellular Specimens}. This is due to edge effects for the finite sized cellular specimens, which have approximately 6 x 6 cells across the footprint area. The way cells have to be truncated to fit within the specimen bounding box varies between the cellular architectures. Recall (see Section \ref{SubSec:Cellular Specimens}) that all cellular geometries were designed with a nominal relative density of 20\%.

For most cellular geometries and process parameter sets, the measured density values are less than the nominal (i.e.\ CAD) values. As observed with the dogbone specimens, process variant 2 leads to higher densities than variants 1 and 3. Process parameter set 2 also produces specimens closest to the nominal density values, across all of the cellular geometries. Comparing the two versions of the octet truss (with 0$^{\circ}$ or 15$^{\circ}$ minimum strut angles), the geometry difference did not lead to a significant difference in density. The differences between the nominal and measured density values, and between process parameter sets might be due to feature resolution at the scale of struts or facets, or porosity at the microstructural scale (as observed with the dogbone specimens). In the following sections, the as-built cellular materials are analysed at different length scales in order to address this question.

\subsection{Cellular Geometry Resolution }
\label{SubSec:Meso-Level Analysis}\vspace{2pt}

In this section, X-ray computed tomography (CT) and scanning electron microscopy (SEM) are used to assess how accurately the cellular geometries have been realised by the three LPBF process parameter sets. The X-ray CT images were made using a Nikon X-TEK XT H 225ST system.

\begin{figure}[htp]
	\begin{subfigure}[!h]{1\textwidth}
		\centering
		\includegraphics[scale=0.54]{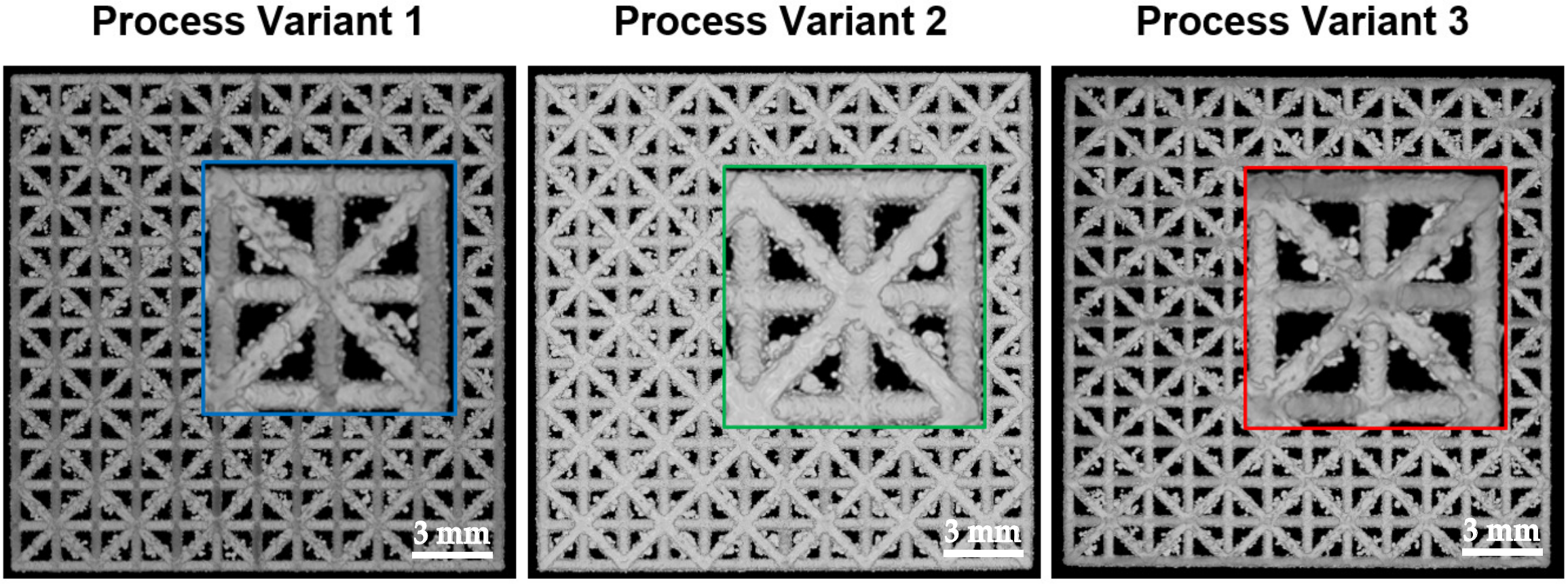}
		\caption{}
		\label{Fig:XRayOctet0}
	\end{subfigure}
	
	\begin{subfigure}[!h]{1\textwidth}
		\centering
		\includegraphics[scale=0.35]{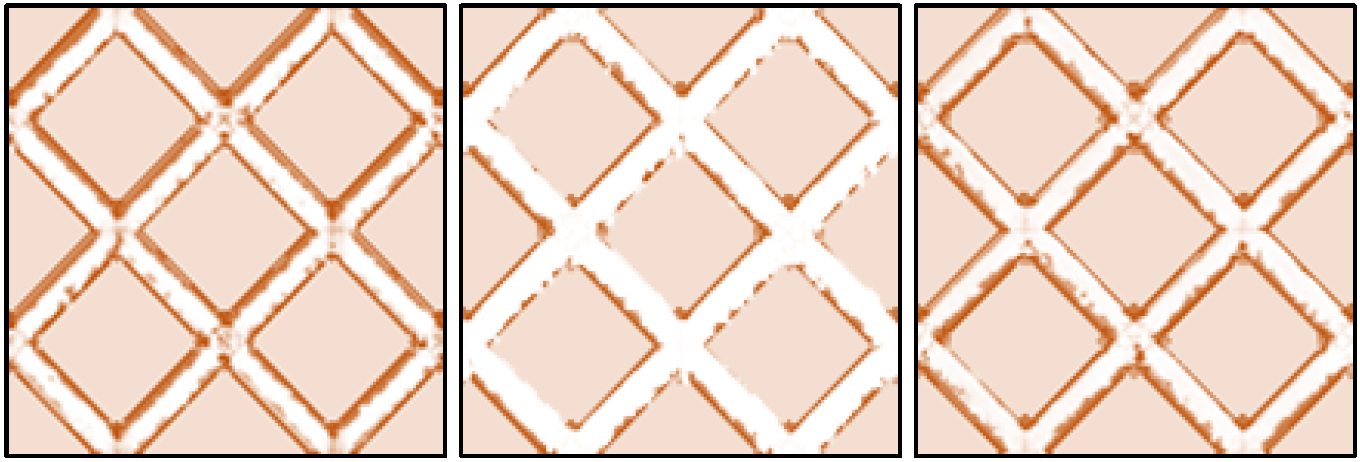}
		\caption{}
		\label{Fig:OverlayOctet0}
	\end{subfigure}
	\caption{(a) X-ray CT scans and (b) overlay of the X-ray CT scan (in white) on the CAD model (in orange) of the octet truss (0$^{\circ}$ version) manufactured with the three process variants.}
\label{Fig:XRayOctetComb}
\end{figure}

\begin{figure}[htp]
	\begin{subfigure}[!h]{1\textwidth}
		\centering
		\includegraphics[scale=0.533]{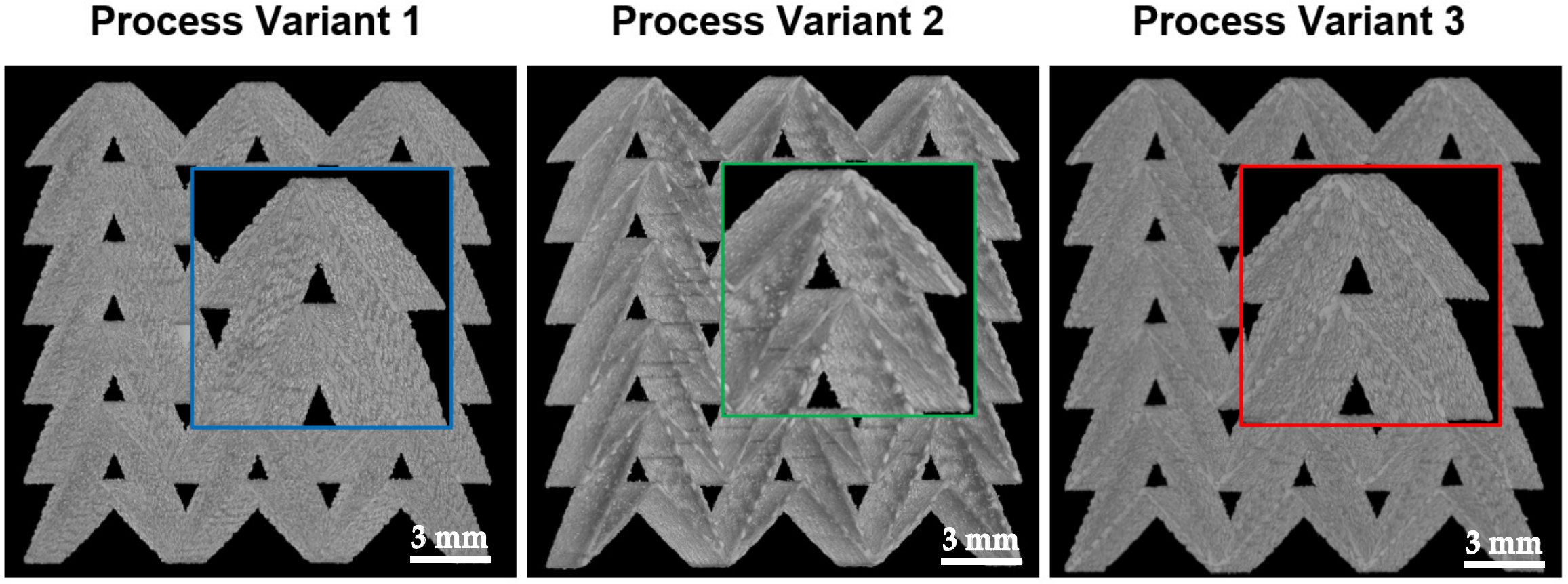}
		\caption{}
		\label{Fig:XRayOrigami}
	\end{subfigure}
	
	\begin{subfigure}[!h]{1\textwidth}
		\centering
		\includegraphics[scale=0.3]{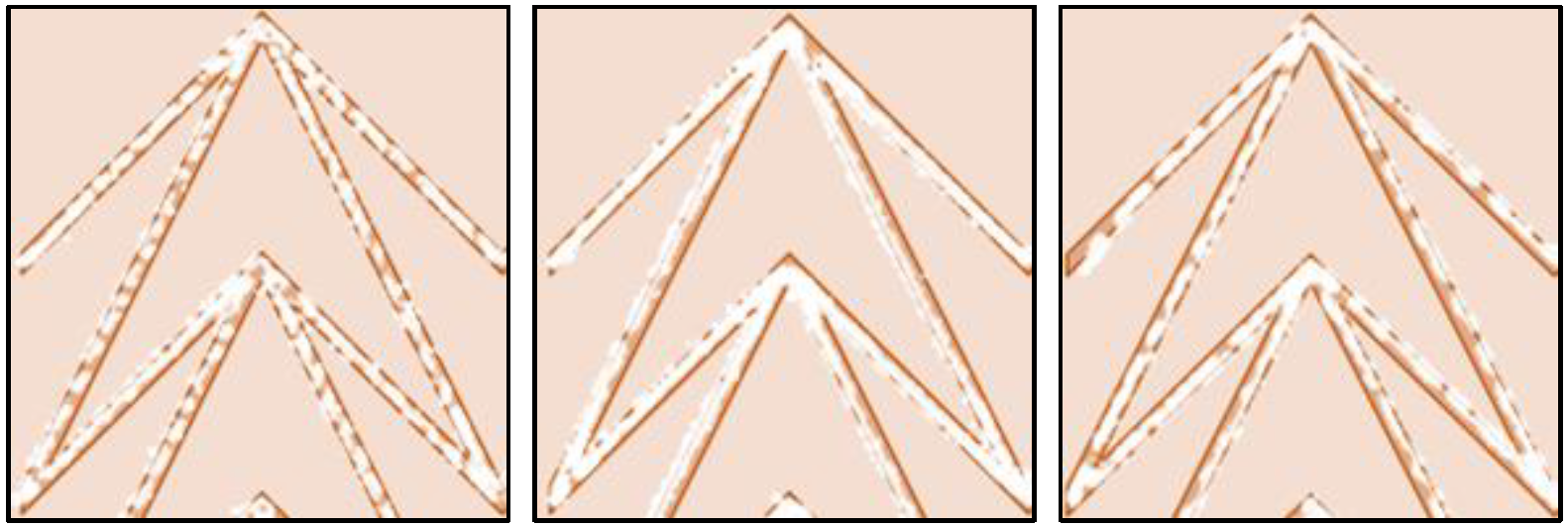}
		\caption{}
		\label{Fig:OverlayOrigami}
	\end{subfigure}
	\caption{(a) X-ray CT scans and (b) overlay of the X-ray CT scan (in white) on the CAD model (in orange) of the stacked Miura-ori origami structure manufactured with the three process variants.}
\label{Fig:XRayOrigamiComb}
\end{figure}

Representative examples of the X-ray CT images are given in Figs. \ref{Fig:XRayOctetComb} and \ref{Fig:XRayOrigamiComb}, showing the octet truss and origami-inspired stacked Miura-ori. In each case, a slice through the 3D X-ray CT scan is overlayed onto the CAD model, to show how accurately the strut and wall thicknesses have been reproduced. For the octet truss, process parameter set 2 shows the closest match, with process variants 1 and 3 appearing to be slightly under-built. For the stacked Miura-ori, the facet thicknesses appear to be similar for the three process variants, and all slightly under-built versus the CAD model. Porosity is also apparent in the X-ray CT slices, and is more pronounced for process parameter sets 1 and 3. The apexes of the Miura-ori structure, where four facets converge with an acute internal angle, are not accurately resolved by the LPBF processes, with some over-building apparent. This may contribute to the measured density of these specimens with process variant 2 exceeding the nominal value (Fig. \ref{Fig:ThreeDensityCellular22Jan20}).

\begin{figure}[H]
	\centering
	\includegraphics[scale=1.0]{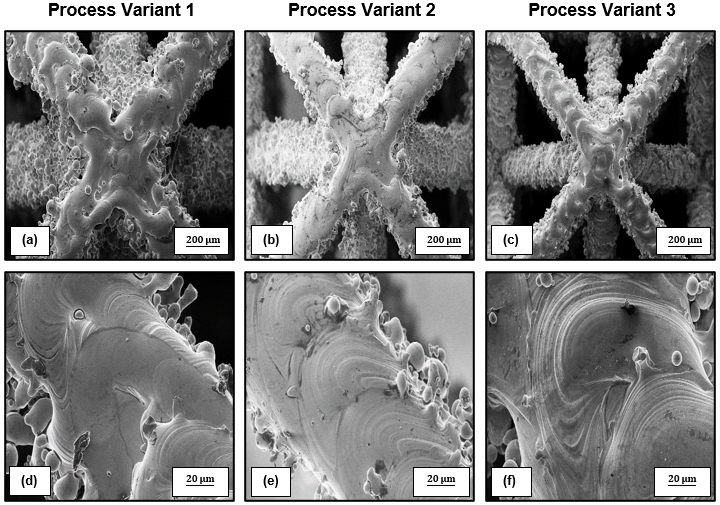}
	\caption{SEM images of the octet truss (0$^{\circ}$ version) manufactured with the three process variants, at two different magnifications.}
	\label{Fig:SEMOctet0}
\end{figure}

\begin{figure}[H]
	\centering
	\includegraphics[scale=1.0]{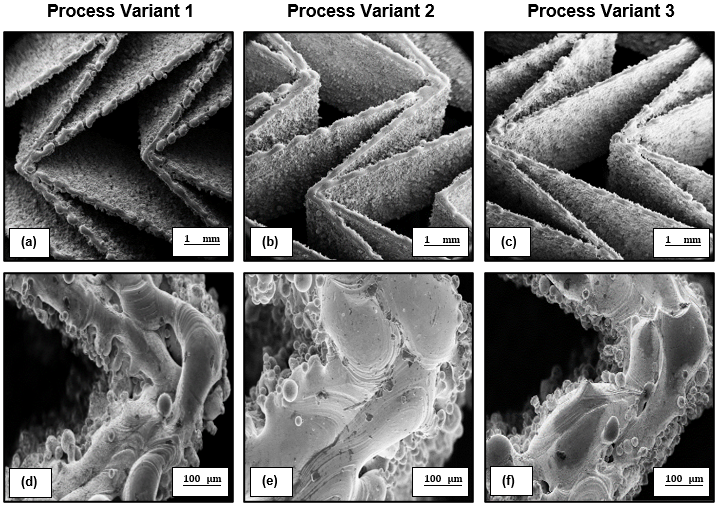}
	\caption{SEM images of the stacked Miura-ori origami structure manufactured with the three process variants, at two different magnifications.}
	\label{Fig:SEMOrigami}
\end{figure}

The X-ray CT scans also show surface roughness. On visual inspection of the specimens, those manufactured with a higher laser power (parameter sets 2 and 3) have a lighter colour compared to process variant 1, which might indicate differences in surface roughness. SEM was used to provide higher magnification images of the surfaces, shown in Figs. \ref{Fig:SEMOctet0} and \ref{Fig:SEMOrigami} for the octet truss and the origami structure. Roughness appears to be partly a result of the laser pulsing strategy, with the distinct melt pools more obvious in these thin-walled structures compared to the dogbones, and partly due to adhered un-melted powder particles. Similar features are apparent for all three process parameter sets. Adhered powder particles are commonly observed for LPBF processing \cite{Kasperovich2021,Zou2022}. As discussed in the first section of the Supplementary Material, a non-zero strut angle has a positive effect on the integrity of the octet samples.

\subsection{Microstructural Characterisation}
\label{SubSec:Micro-Level Analysis}\vspace{2pt}

\begin{figure}[H]
	\begin{subfigure}[!h]{0.49\textwidth}
		\centering
		\includegraphics[scale=0.30]{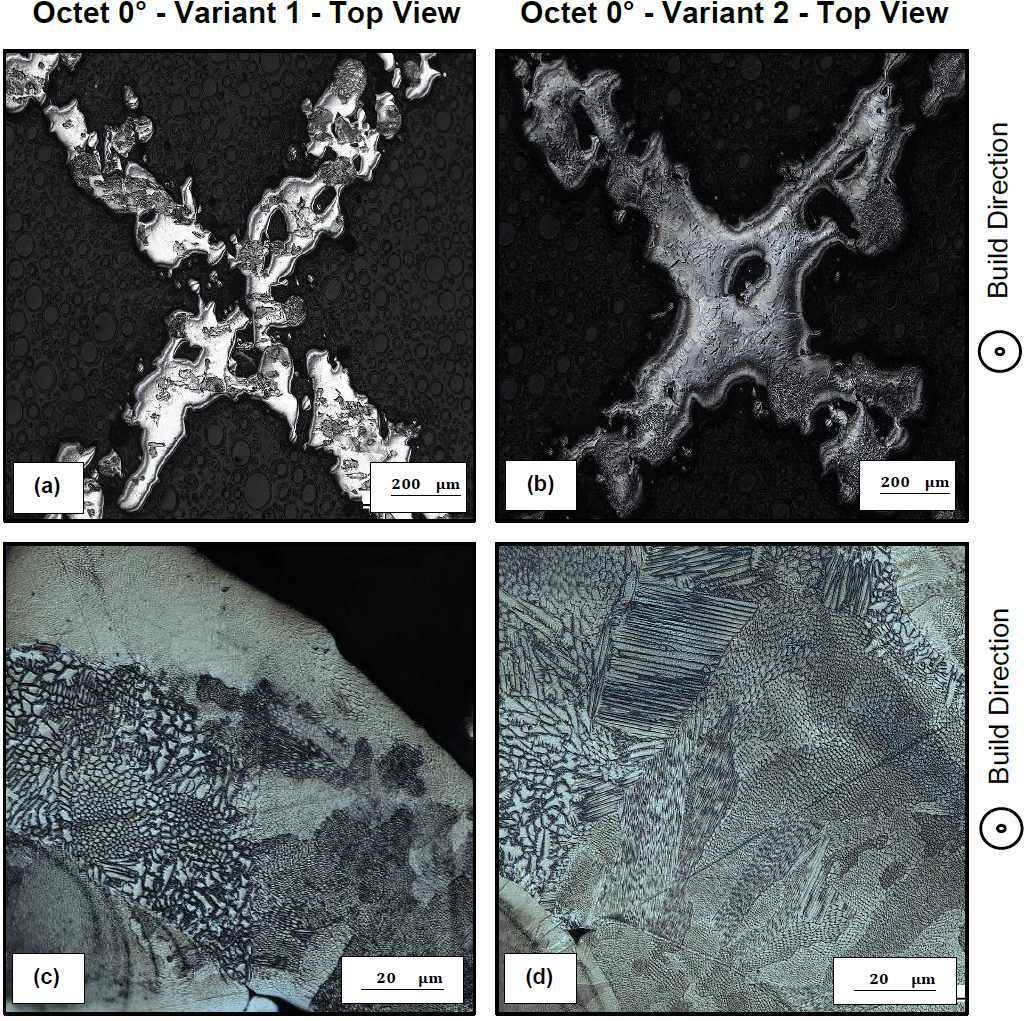}
		\label{Fig:MicrostructureOctet0Top}
	\end{subfigure}
	\begin{subfigure}[h]{0.49\textwidth}
		\centering
		\includegraphics[scale=0.778]{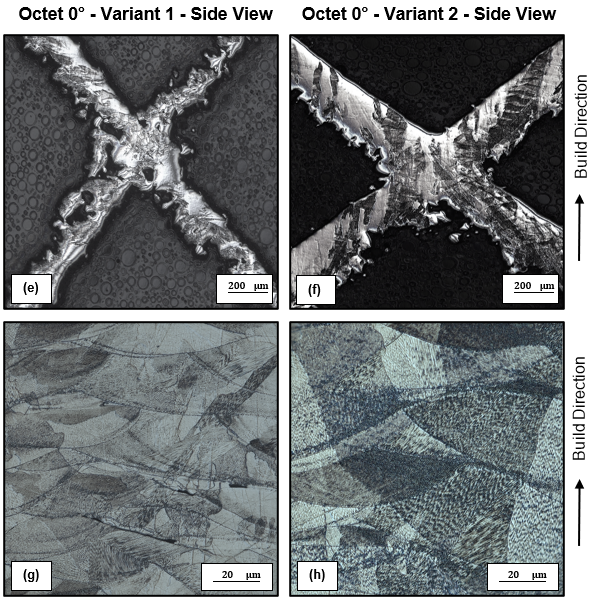}
		\label{Fig:MicrostructureOctet0Side}
	\end{subfigure}
	\caption{Optical microscope images for the octet truss (0$^{\circ}$ version) manufactured with process variants 1 and 2: (a-d) top view at two different magnifications, and (e-h) side view at two different magnifications.}
	\label{Fig:MicrostructureOctet0}
\end{figure}

\begin{figure}[H]
	\begin{subfigure}[h]{0.49\textwidth}
		\centering
		\includegraphics[scale=0.30]{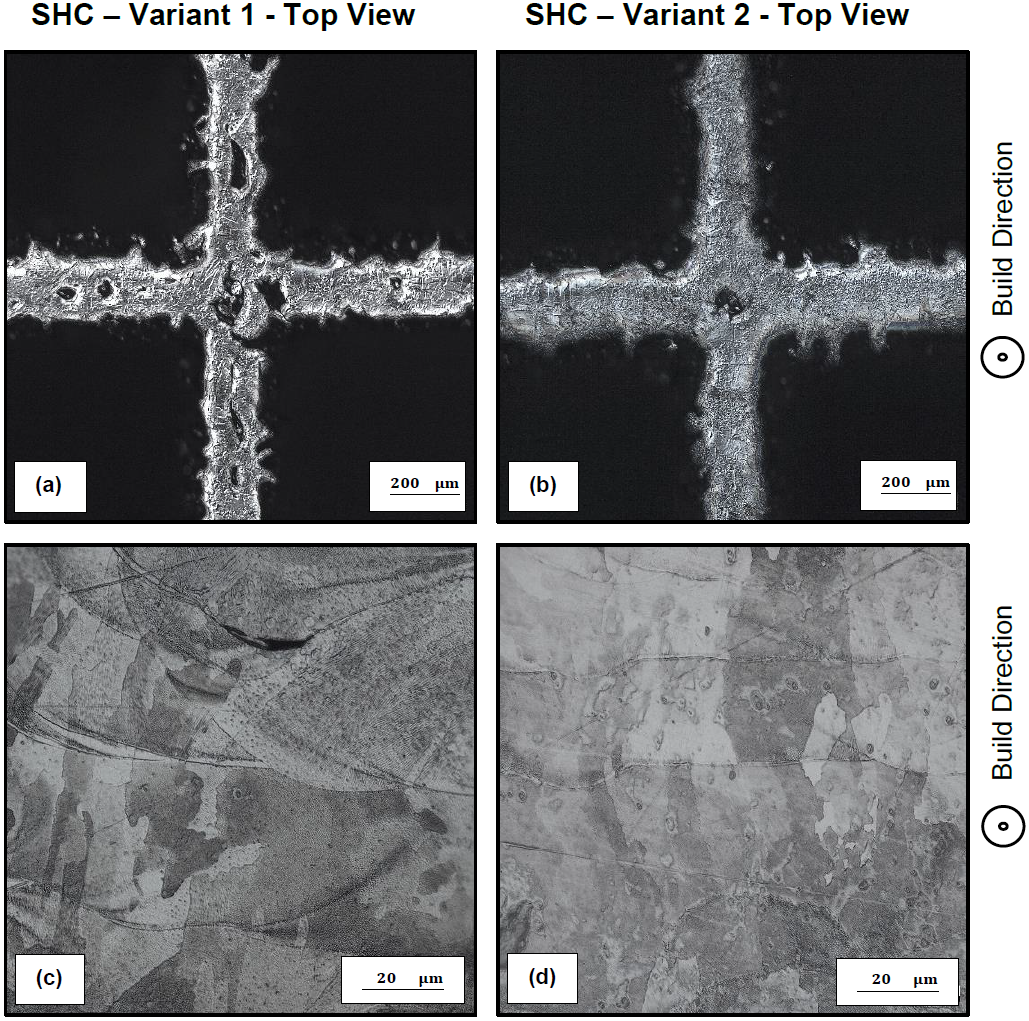}
		\label{Fig:MicrostructureSHCTop}
	\end{subfigure}
	\begin{subfigure}[h]{0.49\textwidth}
		\centering
		\includegraphics[scale=0.30]{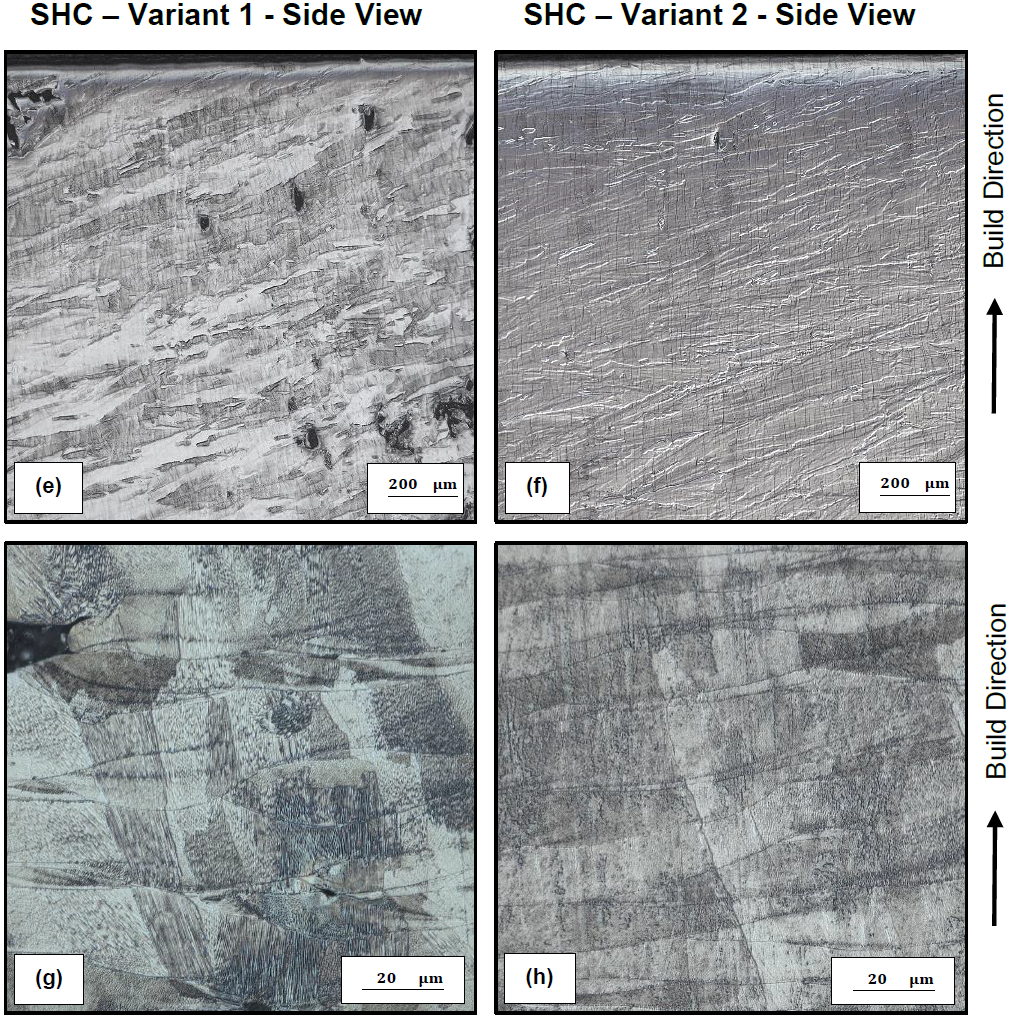}
		\label{Fig:MicrostructureSHCSide}
	\end{subfigure}
	\caption{Optical microscope images for the square honeycomb manufactured with process variants 1 and 2: (a-d) top view at two different magnifications, and (e-h) side view at two different magnifications.}
	\label{Fig:MicrostructureSHC}
\end{figure}

Samples of all five cellular geometries were sectioned, mounted, polished, and etched with Glycerygia (15 ml glycerol, 10 ml HCl, 5 ml HNO$_3$) for 60 - 75 seconds to reveal the microstructure. Both top and side views of the specimens (see Fig. \ref{Fig:FifteenSpecimens_v2}) were observed under dark field optical microscopy at different levels of magnification. Selected micrographs are shown in Figs. \ref{Fig:MicrostructureOctet0} and \ref{Fig:MicrostructureSHC} for the octet truss and the square honeycomb, manufactured using process variants 1 and 2.

At the microstructural level, the cellular materials are very similar to the dogbone tensile specimens described previously in Section \ref{SubSec:Microstructural Characterisation of Solid Dogbones}. There is higher porosity for process parameter set 1, for both the octet truss and the square honeycomb. The distribution, shape and size of the pores is again irregular. At higher magnification, the grain structure, size of the melt pool features and sub-grain structures are similar for the two process variants and cellular geometries.

\section{Quasi-Static Compressive Response}
\label{Sec:Quasi-Static Compressive Response}

The LPBF cellular specimens were compressed quasi-statically in order to measure the strength and energy absorption, and their sensitivity to the process parameters. The loading was parallel to the build direction (see Fig. \ref{Fig:FifteenSpecimens_v2}). All specimens were tested as-manufactured; no heat treatment or surface finishing was used. The specimens were compressed at a nominal strain rate of 10$^{-3}$ s$^{-1}$ using an Instron screw-driven materials testing machine.  A laser extensometer was used to measure the platen displacement, and a clip gauge was used to check the small strain measurements. The nominal strain was calculated by dividing the platen displacement by the original height of the cellular specimens ($H$ = 12 mm). The nominal stress was calculated by dividing the force obtained from the test machine load cell by the total footprint area (see Fig. \ref{Fig:UnitCells}).

\begin{figure}[h]
	\begin{subfigure}[h]{0.34\textwidth}
		\centering
		\includegraphics[width=0.85\linewidth]{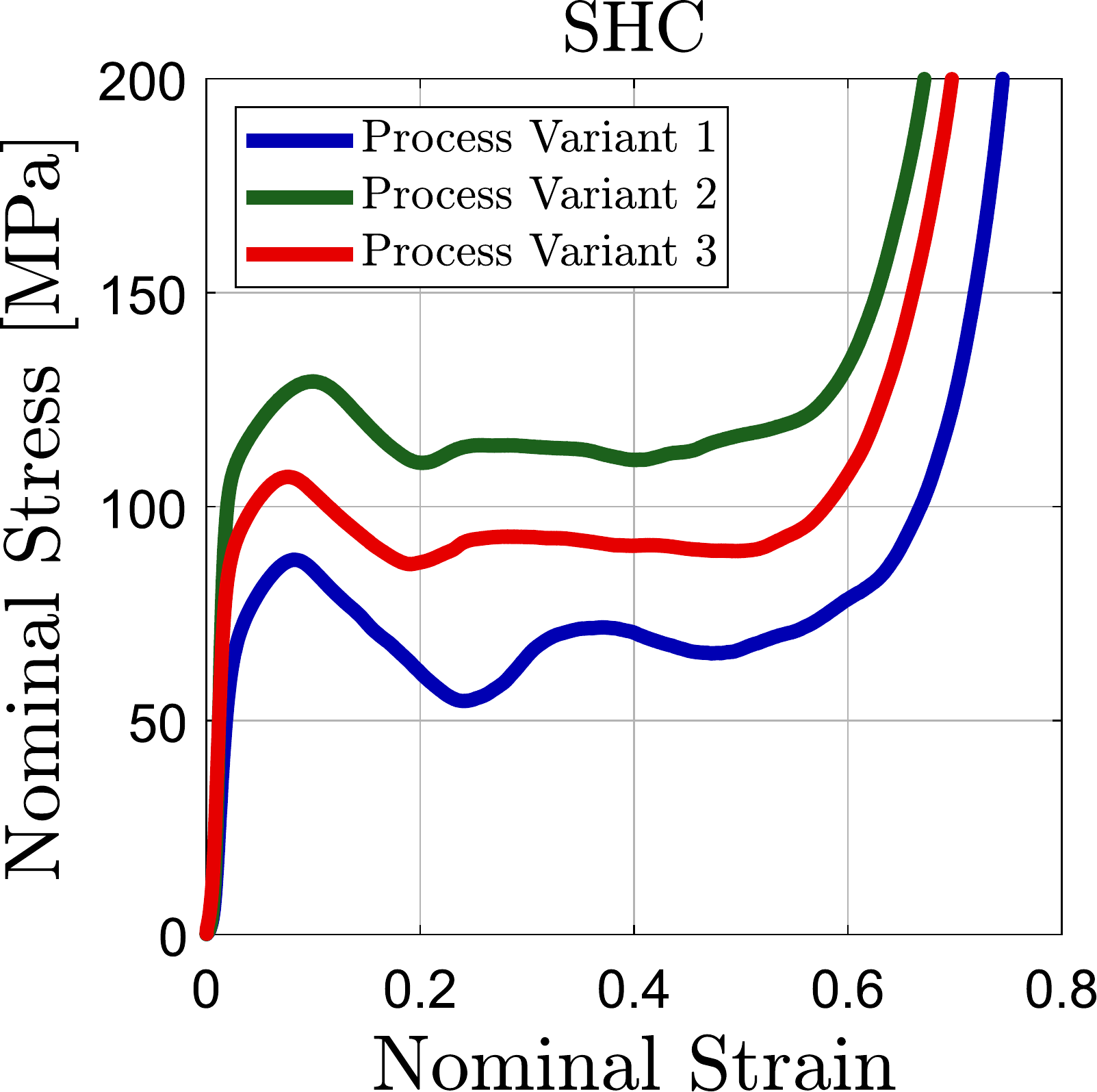}
		\caption{}
		\label{Fig:SHCNomNom}
	\end{subfigure}%
	\begin{subfigure}[h]{0.34\textwidth}
		\centering
		\includegraphics[width=0.85\linewidth]{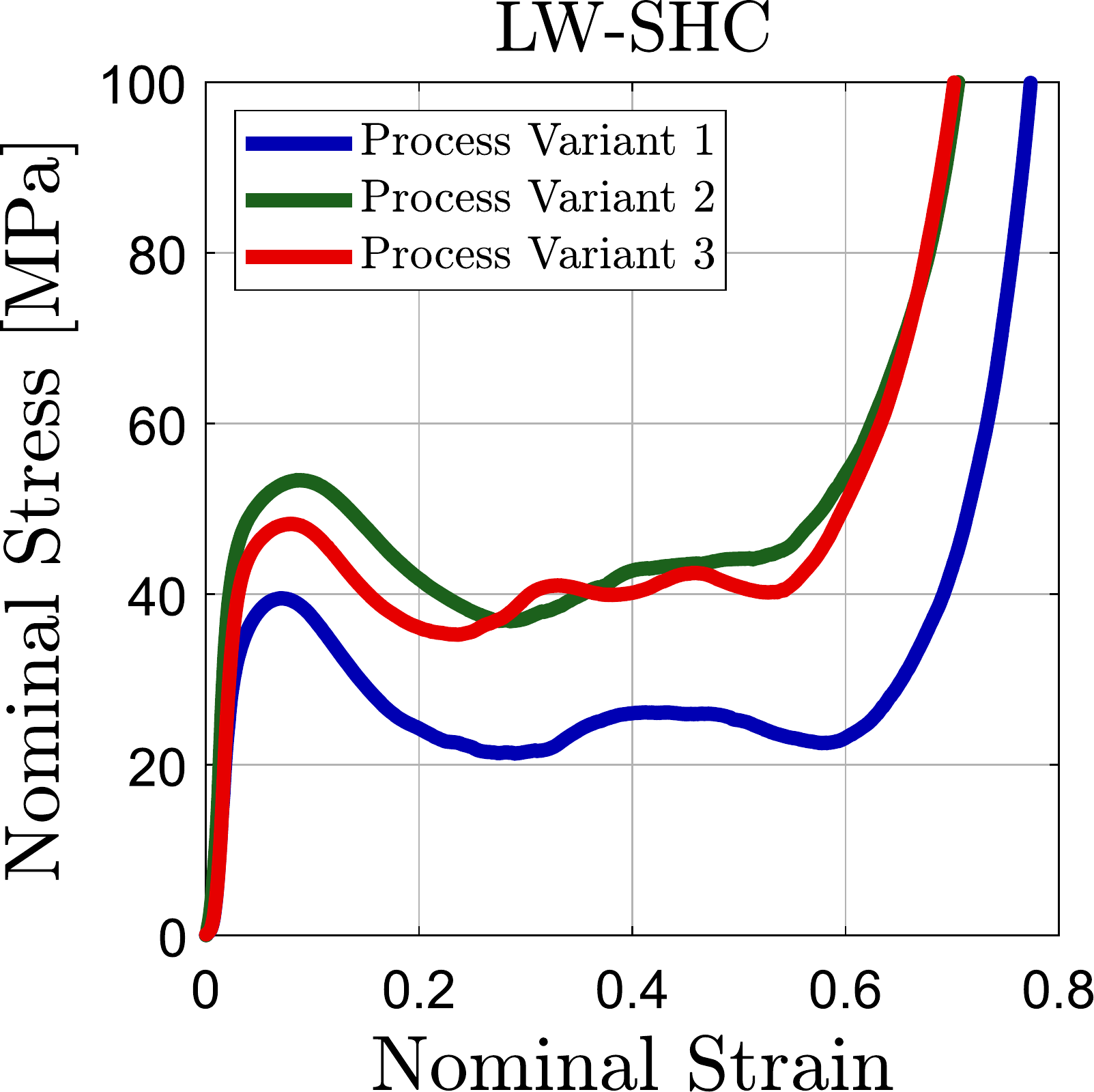}
		\caption{}
		\label{Fig:LWSHCNomNom}
	\end{subfigure}
	\begin{subfigure}[h]{0.34\textwidth}
		\centering
		\includegraphics[width=0.85\linewidth]{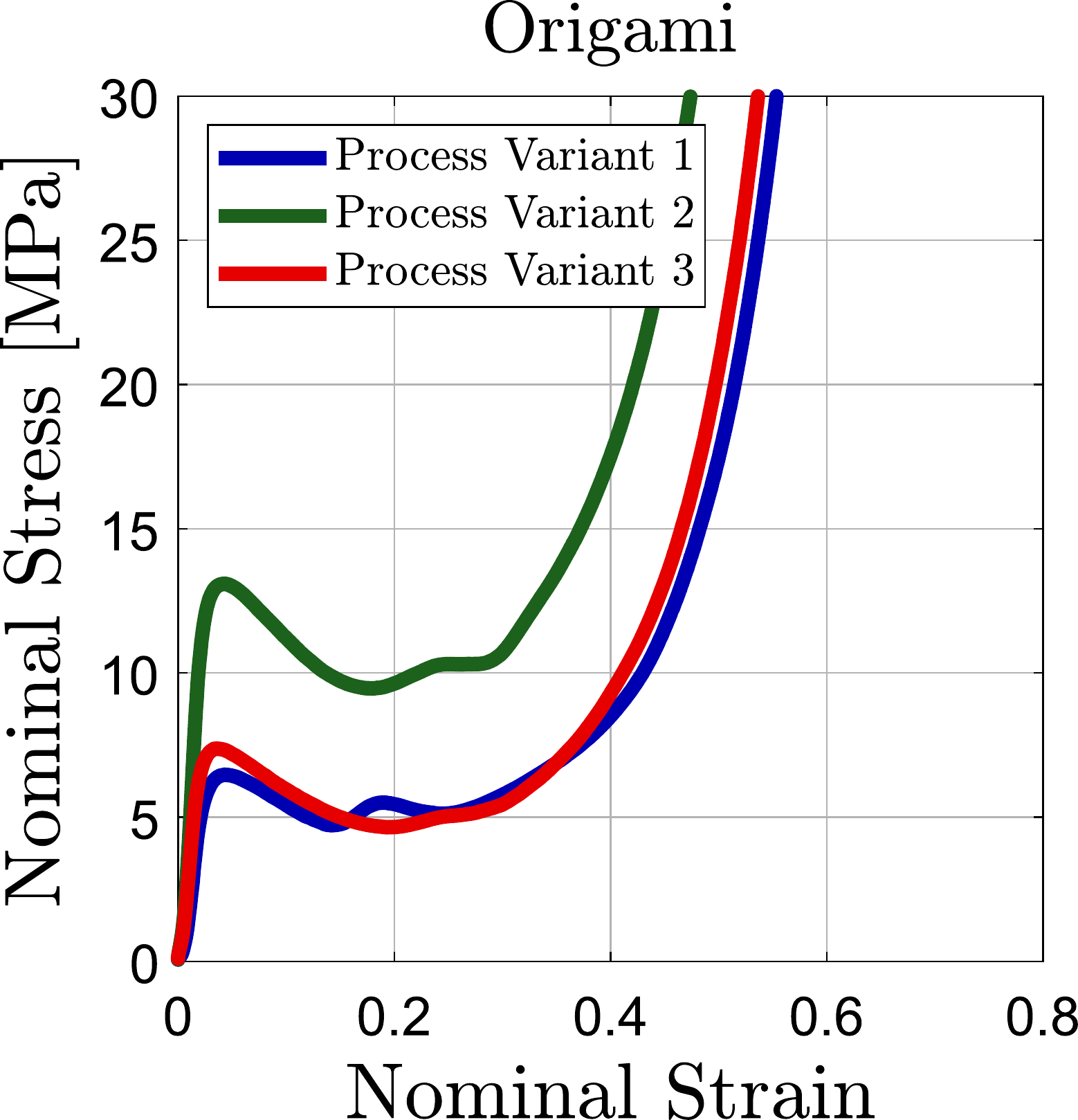}
		\caption{}
		\label{Fig:OrigamiNomNom}
	\end{subfigure}
	
\begin{subfigure}[h]{0.5\textwidth}
		\centering
		\includegraphics[width=0.57\linewidth]{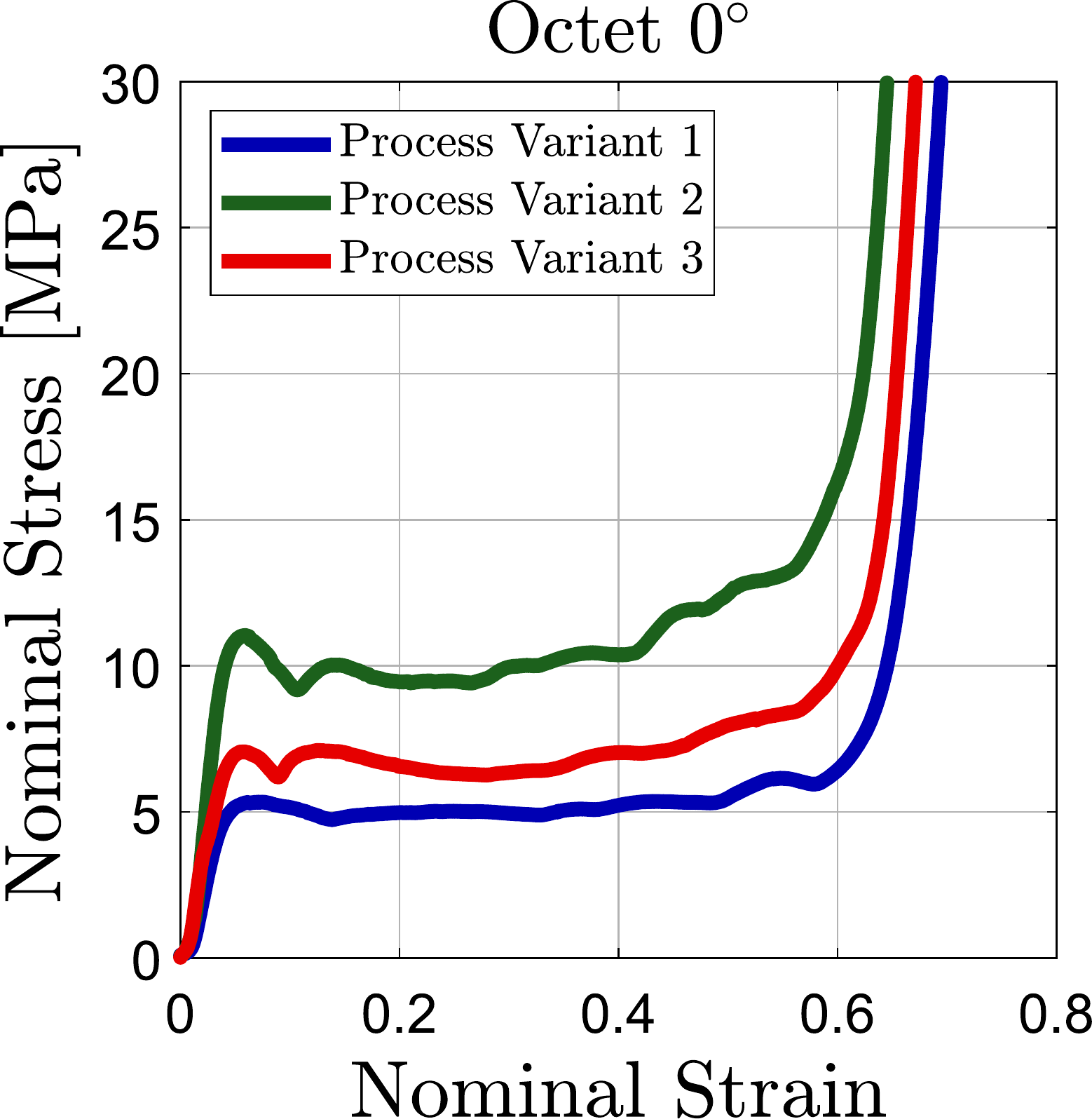}
		\caption{}
		\label{Fig:Octet0NomNom}
	\end{subfigure}%
	\begin{subfigure}[h]{0.5\textwidth}
		\centering
		\includegraphics[width=0.57\linewidth]{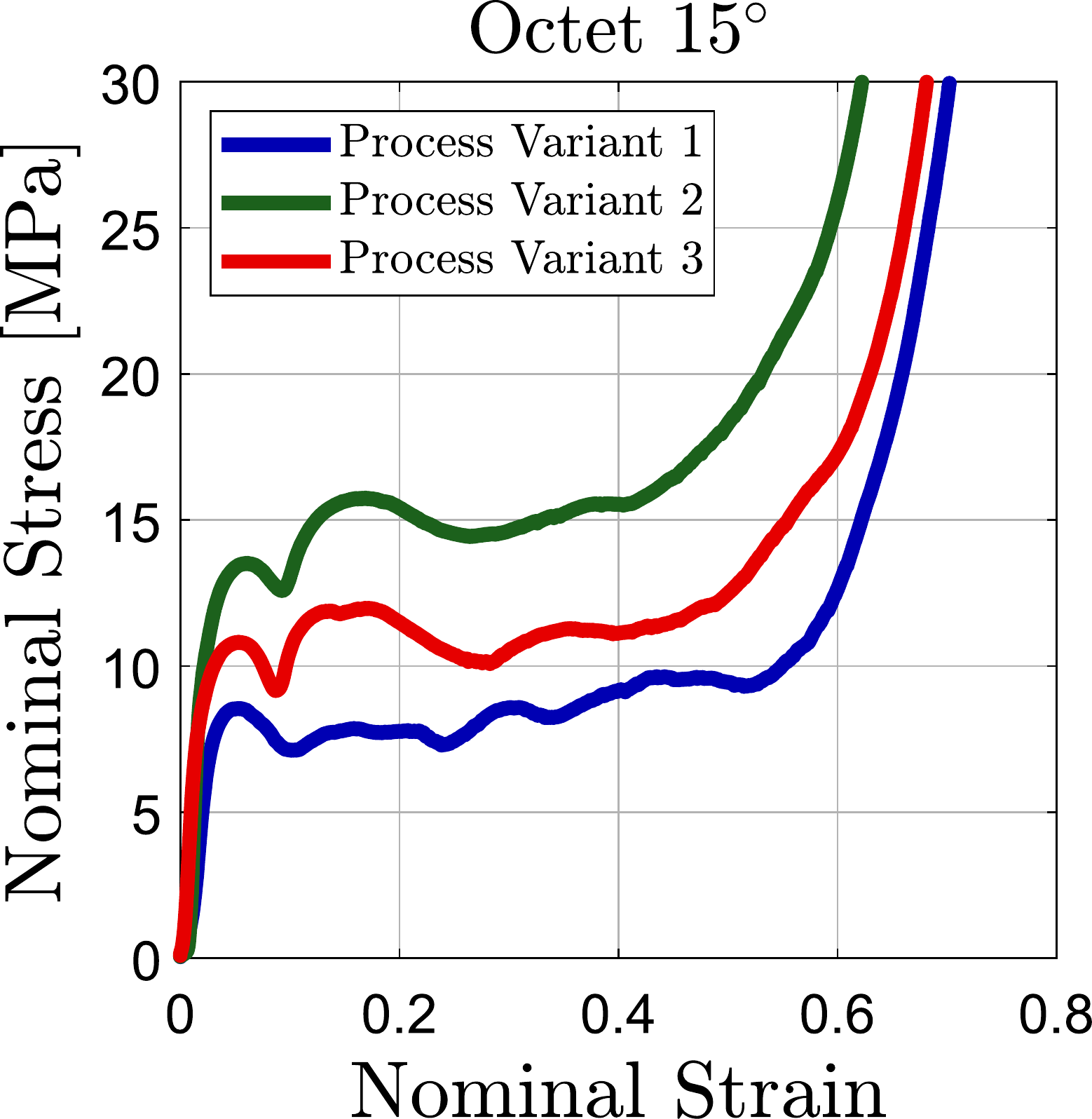}
		\caption{}
		\label{Fig:Octet15NomNom}
	\end{subfigure}
	\caption{Nominal stress versus strain of the five geometries as-manufactured with the three process variants: (a) SHC, (b) LW-SHC, (c) Origami, (d) Octet Lattice 0$^{\circ}$, (e) Octet Lattice 15$^{\circ}$.}
	\label{Fig:StressStrainEachGeometry}
\end{figure}

The nominal stress versus nominal strain results for each cellular geometry and process variant are given in Fig. \ref{Fig:StressStrainEachGeometry}. In all cases, a peak in strength is reached, followed by softening due to the onset of buckling of struts and cell walls. A plateau in strength follows as the cells progressively collapse. Finally, the stress rises as the cellular material begins to densify. The origami structure has the lowest densification strain. This was also observed in the experiments of Harris and McShane \cite{Harris2020}, and attributed to the interactions between the facets as the origami specimen folds during compression. The origami structure is also auxetic, due to the folding kinematics of the facets. The contraction transverse to the direction of loading contributes to this early densification \cite{Harris2020}. The square honeycomb geometry has the highest strength, roughly an order of magnitude larger than the octet truss and the origami structure. The lattice-walled square honeycomb has a strength in between these cases. All cellular geometries show a similar sensitivity to the LPBF process parameter sets. The highest peak strength is seen for process variant 2 (intermediate power and exposure time), and the lowest for process variant 1 (lowest power and longest exposure time).

\subsection{Performance Metrics}
\label{SubSec:Performance Metrics}\vspace{2pt}

Next, a number of performance metrics relevant to energy absorption applications are evaluated from the stress-strain curves. The peak stress $\sigma_{p}(\varepsilon_1)$ is the maximum nominal stress observed up to a nominal strain $\varepsilon_1$. The volumetric energy absorption up to strain $\varepsilon_1$ is
\begin{equation}\label{Eq:EnergyAbsorption}
W(\varepsilon_1)=\int_{0}^{\varepsilon_1} \sigma \; d\varepsilon
\end{equation}
\noindent where $\sigma$ and $\varepsilon$ are the nominal stress and strain. Following Tan et al. \cite{Tan2002}, the energy absorption efficiency ($\mu$) for compression up to $\varepsilon_1$ is given by
\begin{equation}\label{Eq:Efficiency}
\mu(\varepsilon_1)=\frac{W(\varepsilon_1)}{\sigma_p(\varepsilon_1)}
\end{equation}

In this section, these performance metrics are evaluated and compared up to the densification strain: $\varepsilon_1 = \varepsilon_D$.  The densification strain $\varepsilon_D$ is defined to be the point of maximum efficiency $\mu(\varepsilon_1)$ \cite{Tan2002}. The results are presented in Table \ref{Tab:Metrics}. Some publications choose different strain ranges over which to calculate these metrics. Therefore, in the Supplementary Material, Tables 1 and 2 give these metrics with the alternative choices $\varepsilon_1$=0.3 and $\varepsilon_1$=0.5.

The efficiency $\mu$ is an indicator of the shape of the compressive stress-strain curve: softening or strain hardening after the initial peak in stress both diminish $\mu$. The square honeycomb, lattice-walled honeycomb and octet truss specimens have similar efficiencies, falling in the range $\mu \approx $ 0.4 - 0.6. The origami structure has a lower efficiency, as the compressive stress-strain curve shows only a short plateau phase before densification begins. There is no strong dependence of $\mu$ on the process parameter set/variant. As shown (e.g.) in Fig. \ref{Fig:StressStrainEachGeometry}(a) the process parameters have a notable influence on the peak stress $\sigma_p$ and the volumetric energy absorption $W$ (area under the curve), but both increase (or decrease) to a similar extent and thus the impact on the efficiency parameter $\mu$ is minimal - see Eq. (\ref{Eq:Efficiency}).

\begin{table}[H]
	\centering
	\caption{Performance metrics evaluated up to the densification strain, for the five cellular geometries manufactured with each process parameter set.}
	\label{Tab:Metrics}
		\begin{tabular}{|c|c|c|c|c|c|c|}
			\hline
			\begin{tabular}[c]{@{}c@{}}Performance\\ Metric\end{tabular} & \begin{tabular}[c]{@{}c@{}}Parameter\\ Set\end{tabular} & SHC   & LW-SHC & Origami & Octet $0^\circ$ & Octet $15^\circ$ \\ \hline
			& {\color[HTML]{3531FF} \textbf{1}} & 87.7  & 39.8   & 6.4     & 5.3             & 8.5              \\ \cline{2-7}
			& {\color[HTML]{009901} \textbf{2}}                         & 129.8 & 53.7   & 13.1    & 11.0            & 13.4             \\ \cline{2-7}
			\multirow{-3}{*}{$\sigma_{p}$ [MPa]}                               & {\color[HTML]{FE0000} \textbf{3}}                         & 107.0 & 48.6   & 7.4     & 7.2             & 10.8             \\ \hline
			& {\color[HTML]{3531FF} \textbf{1}}                         & 0.64   & 0.69    & 0.33     & 0.59             & 0.54              \\ \cline{2-7}
			& {\color[HTML]{009901} \textbf{2}}                         & 0.59   & 0.60   & 0.34     & 0.56             & 0.49              \\ \cline{2-7}
			\multirow{-3}{*}{$\varepsilon_D$}                            & {\color[HTML]{FE0000} \textbf{3}}                         & 0.60   & 0.59    & 0.36     & 0.56    & 0.49              \\ \hline
			& {\color[HTML]{3531FF} \textbf{1}}                         & 54.4  & 26.4   & 2.1     & 3.1             & 4.6              \\ \cline{2-7}
			& {\color[HTML]{009901} \textbf{2}}                         & 74.0  & 30.8   & 4.3     & 6.1             & 7.4              \\ \cline{2-7}
			\multirow{-3}{*}{$W$($\varepsilon_D$) [MJ/m$^3$]}                         & {\color[HTML]{FE0000} \textbf{3}}                         & 62.3  & 27.4   & 2.6     & 3.9             & 5.6              \\ \hline
			& {\color[HTML]{3531FF} \textbf{1}}                         & 0.62   & 0.67    & 0.32     & 0.50             & 0.48              \\ \cline{2-7}
			& {\color[HTML]{009901} \textbf{2}}                         & 0.57   & 0.58    & 0.33     & 0.46             & 0.41              \\ \cline{2-7}
			\multirow{-3}{*}{$\mu$($\varepsilon_D$)}                                      & {\color[HTML]{FE0000} \textbf{3}}                         & 0.58   & 0.57    & 0.35     & 0.46            & 0.46              \\ \hline
		\end{tabular}
\end{table}

\begin{figure}[H]
	\begin{subfigure}[!h]{0.49\textwidth}
		\centering
	\includegraphics[scale=0.62]{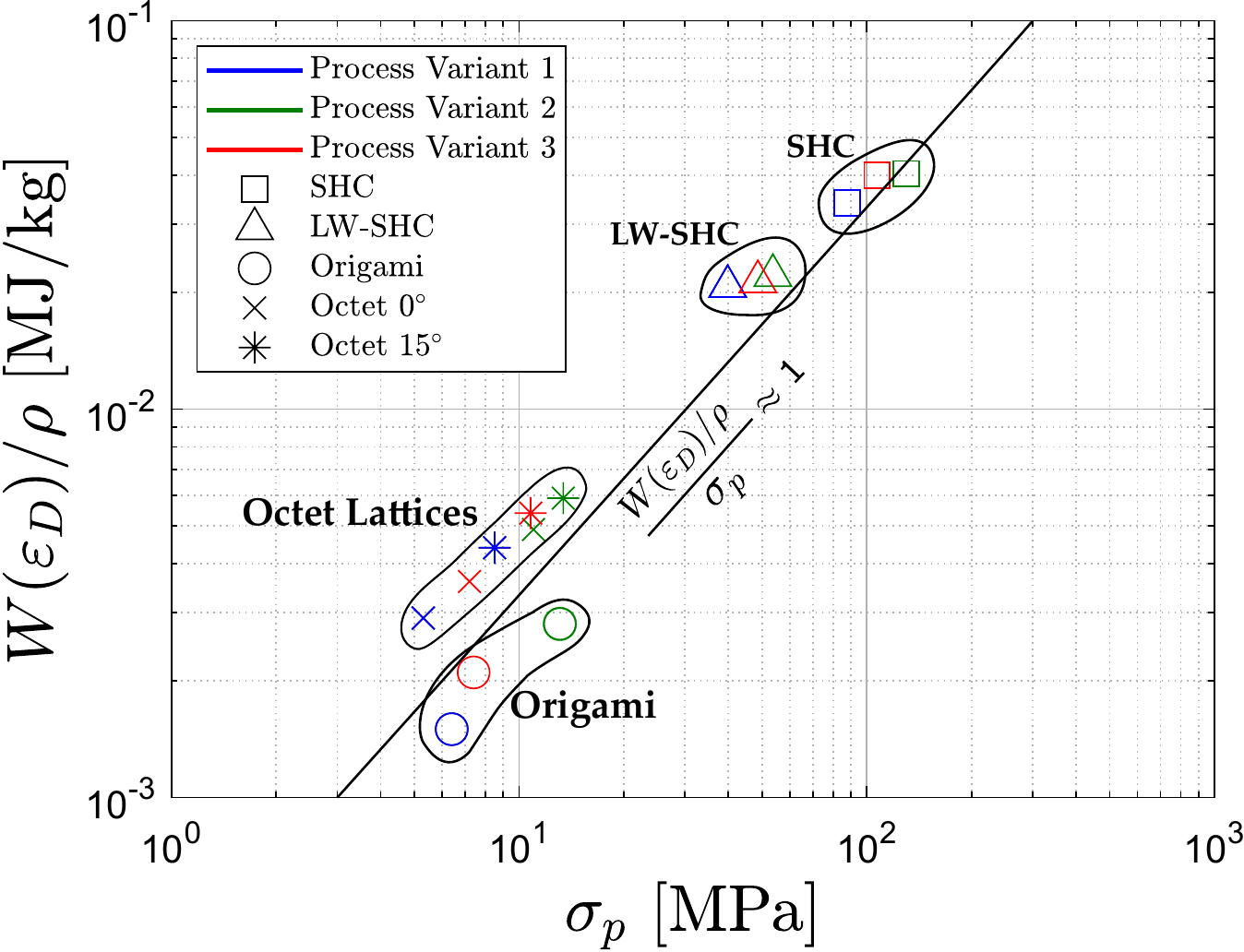}
				\caption{}
		\label{Fig:AshbyLogScaleEnergyAbsorptionbyDensityLogScalePeakStress}
	\end{subfigure}
	\begin{subfigure}[h]{0.49\textwidth}
		\centering
	\includegraphics[scale=0.62]{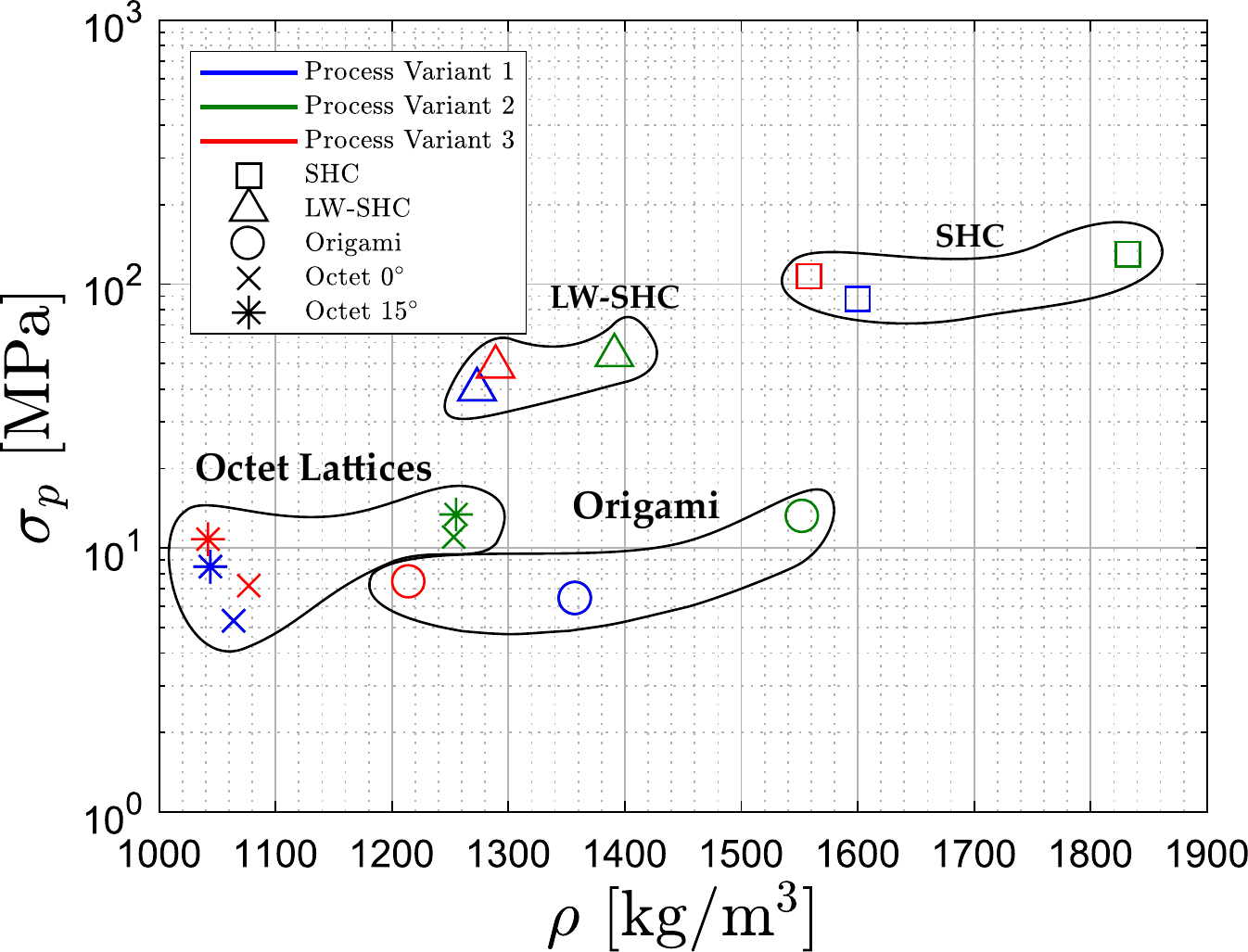}
				\caption{}
		\label{Fig:AshbyLogScalePeakStressLinearAxisDensity}
	\end{subfigure}
	\caption{Peak stress versus (a) energy absorption and (b) density for the five cellular geometries manufactured with the three process variants.}
	\label{Fig:AshbyLogScalePeakStressDensityAbsorption}
\end{figure}

The relationship between energy absorption and peak stress is shown in Fig. \ref{Fig:AshbyLogScalePeakStressDensityAbsorption}(a). The square honeycomb has the highest strength and energy absorption, and the octet truss and origami structure the least. The lattice-walled honeycomb lies in between. The process parameter variants with higher porosities have lower strength and energy absorption compared to process variant 2. However, the effect of sub-optimal processing parameters is (in most cases) much less than the sensitivity to the choice of cellular geometry. The space on the property chart occupied by each cellular geometry remains distinct across all process parameter sets. With the exception of the octet truss and origami structures, whose performance metrics are similar, a `well made' example of one cellular architecture does not move it into the property space of a different design.

The relationship between peak strength and measured density is shown in Fig. \ref{Fig:AshbyLogScalePeakStressDensityAbsorption}(b). Because of the similar efficiencies of these cellular materials, the variation in energy absorption with density is similar, so is omitted here. The conclusions here are similar: the regions of this property space occupied by each cellular geometry are largely distinct as the process is varied. Again, perhaps with the exception of the octet lattice and origami structure, whose properties overlap, the choice of geometry is the most influential design parameter.

\section{Dynamic Compressive Response}
\label{Sec:Dynamic Compressive Response}

In this section, the dynamic compressive behaviour of the LPBF cellular materials is discussed. In the dynamic tests we focus attention on specimens manufactured with process parameter set/variant 2, the low porosity case, with the aim of assessing primarily the influence of loading rate.

\begin{figure}[H]
	\centering
	\includegraphics[scale=0.8]{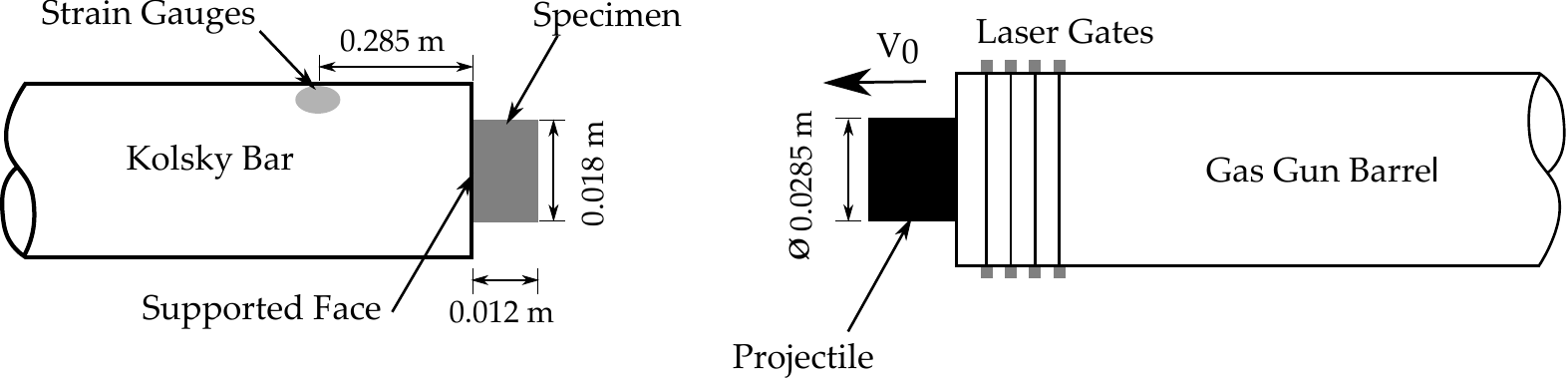}
	\caption{Kolsky bar schematic figure with the supported face configuration.}
	\label{Fig:KolskyBar}
\end{figure}

The dynamic compression tests were performed using a direct impact Kolsky bar, as illustrated in Fig. \ref{Fig:KolskyBar}. The instrumented Kolsky bar has length 2.18 m and diameter 28.5 mm, and is fabricated from maraging steel. The cellular specimens were mounted to the Kolsky bar and impacted by a circular cylindrical steel projectile with mass 0.2 kg and diameter 28.5 mm. The transmitted stress from the distal face of the specimen was measured using a pair of strain gauges located approximately ten bar diameters from the impacted end of the Kolsky bar. The strain gauges were mounted diametrically opposite each other, to eliminate bar bending effects. The tests were performed at an impact velocity of 100 m/s. At this speed, a difference in the local stress at the impacted and distal (i.e.\ supported) faces of the specimens would be expected. Here, we only measure the distal face response; measurements of the stress difference versus the impacted face for LPBF cellular materials with similar geometries can be found in Ozdemir et al. \cite{Ozdemir2015} (lattice materials), Harris et al. \cite{Harris2017} (the square honeycomb, and lattice-walled honeycomb) and Harris and McShane \cite{Harris2021} (the origami structure).

The deformation of the specimen was measured using high speed photography. The position of the leading edges of the projectile and Kolsky bar were tracked in consecutive frames. The specimen nominal strain is calculated by dividing the relative displacement of these leading edges by the undeformed specimen height $H$. As non-uniform straining between the impacted and distal faces of the specimen is possible at higher strain rates, we note that this represents a measure of average deformation along the specimen height. The measured nominal strain versus time is synchronised with the nominal stress versus time obtained from the Kolksy bar, to produce a nominal stress-strain curve.

Projectile velocity-time plots for each impact are given in the Supplementary Material for reference. These show that, for most specimens, the nominal strain rate does not reduce significantly (due to projectile deceleration) during the first 50\% nominal strain. The higher strength of the square honeycomb leads to greater projectile deceleration, and reduces the average strain rate by about 10\% compared to the other specimens.

\begin{figure}[H]
	\centering
	\includegraphics[scale=0.44]{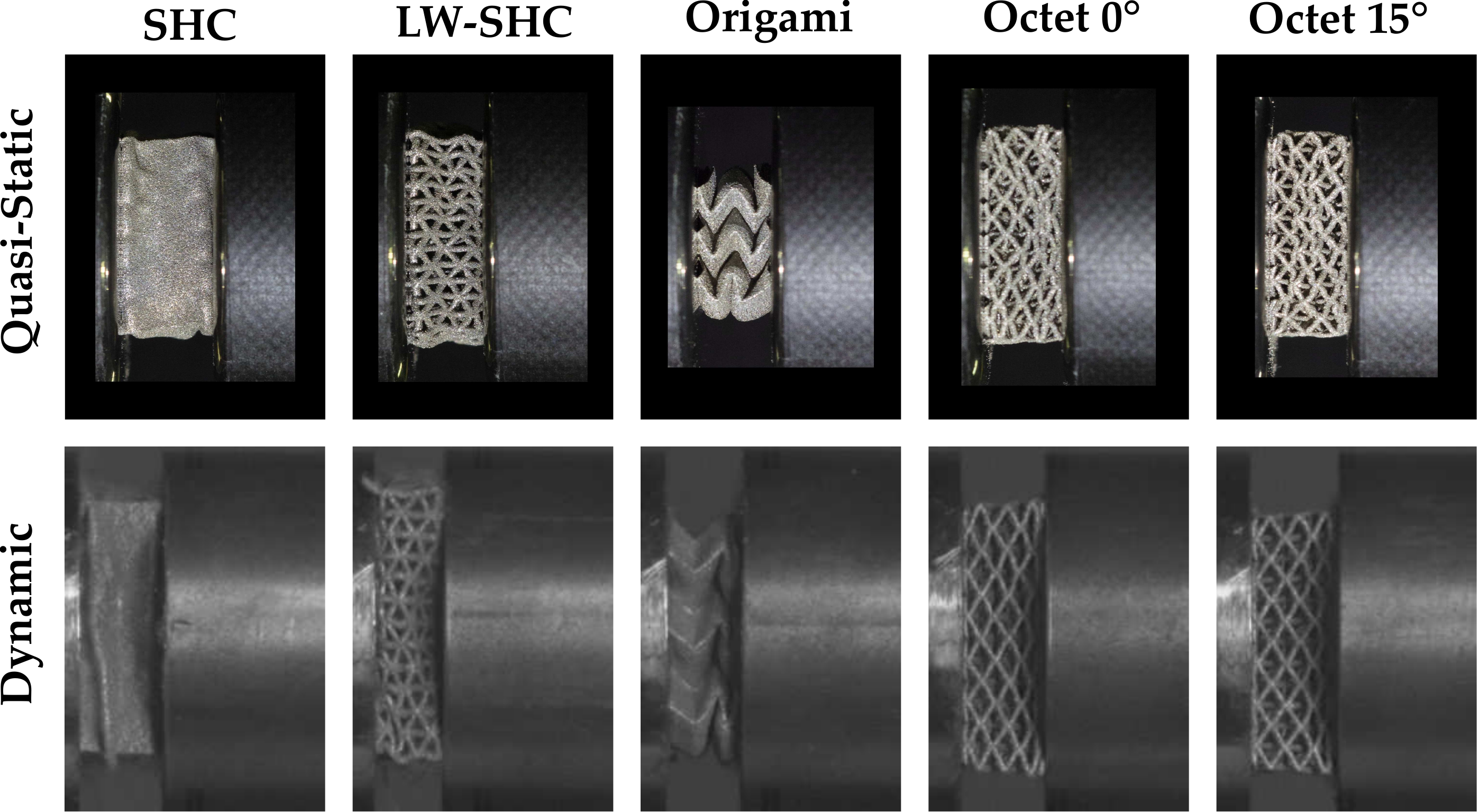}
	\caption{Photographs of the five cellular geometries fabricated with process variant 2 at 50\% nominal strain, during quasi-static and dynamic compression testing. In the dynamic cases, the projectile is impacting from right to left.}
	\label{Fig:QSvsDYN}
\end{figure}

Fig. \ref{Fig:QSvsDYN} compares images of the cellular specimens (process variant 2) from the quasi-static and impact experiments, all compressed to 50\% nominal strain. For the octet lattices, there is less evidence of strut buckling in the dynamic load case. This may be due to inertial stabilisation effects. For the origami structure, the deformation is more concentrated at the impacted face during dynamic compression. The numerical study of Harris and McShane \cite{Harris2021} indicates this is related to wave propagation effects. It also leads to a reduction in the auxetic lateral contraction, increasing the densification strain \cite{Harris2021}. For the square honeycomb, the quasi-static and dynamic modes of collapse are more similar. The higher stiffness and strength of this geometry will reduce wave propagation effects at this impact speed.

\begin{figure}[H]
	\begin{subfigure}{0.5\textwidth}
		\centering
		\includegraphics[width=0.8\linewidth]{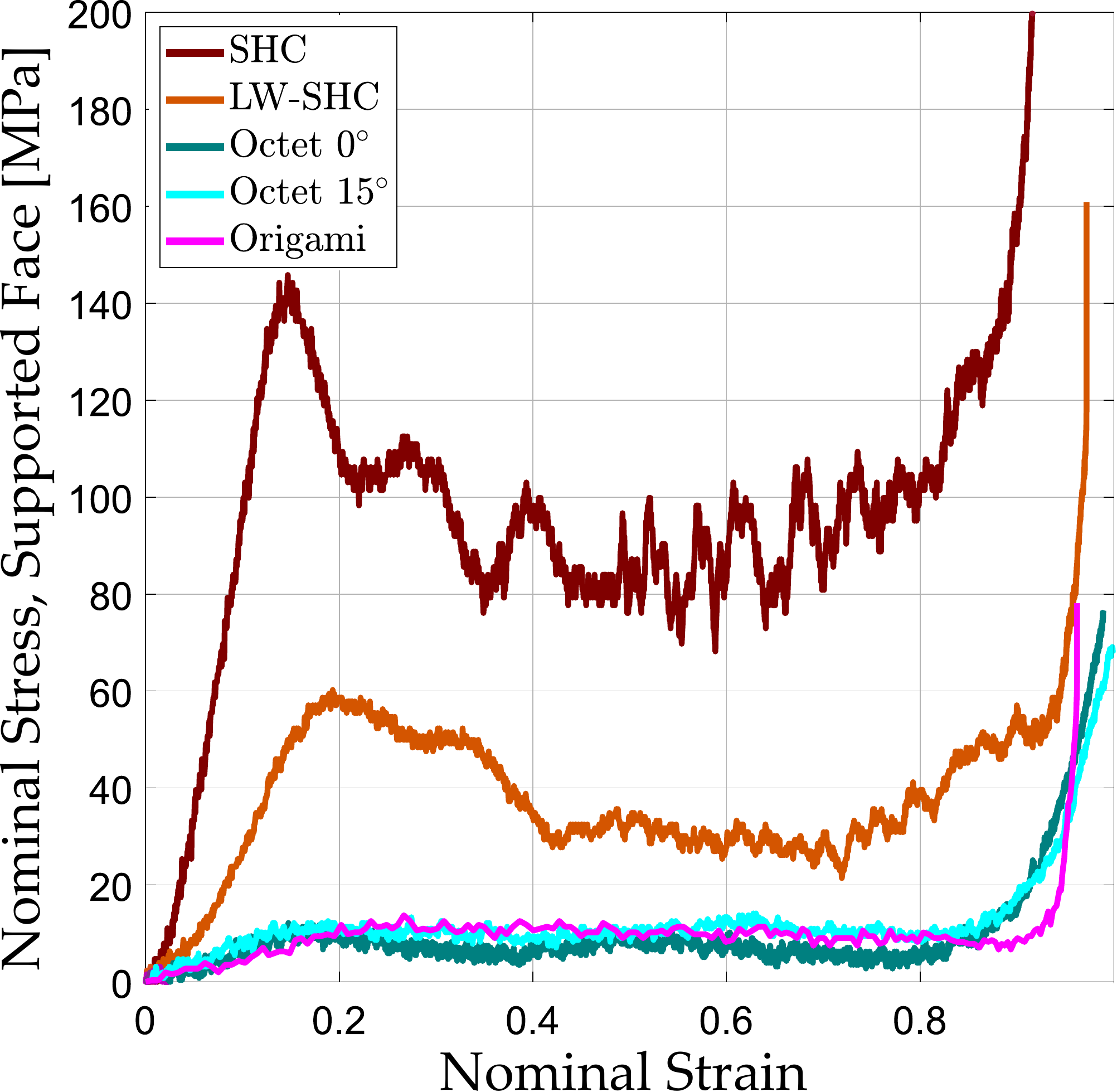}
		\caption{}
		\label{Fig:NominalStressNominalStrain5geometries125W100ms_v2}
	\end{subfigure}%
	\begin{subfigure}{0.5\textwidth}
		\centering
		\includegraphics[width=0.8\linewidth]{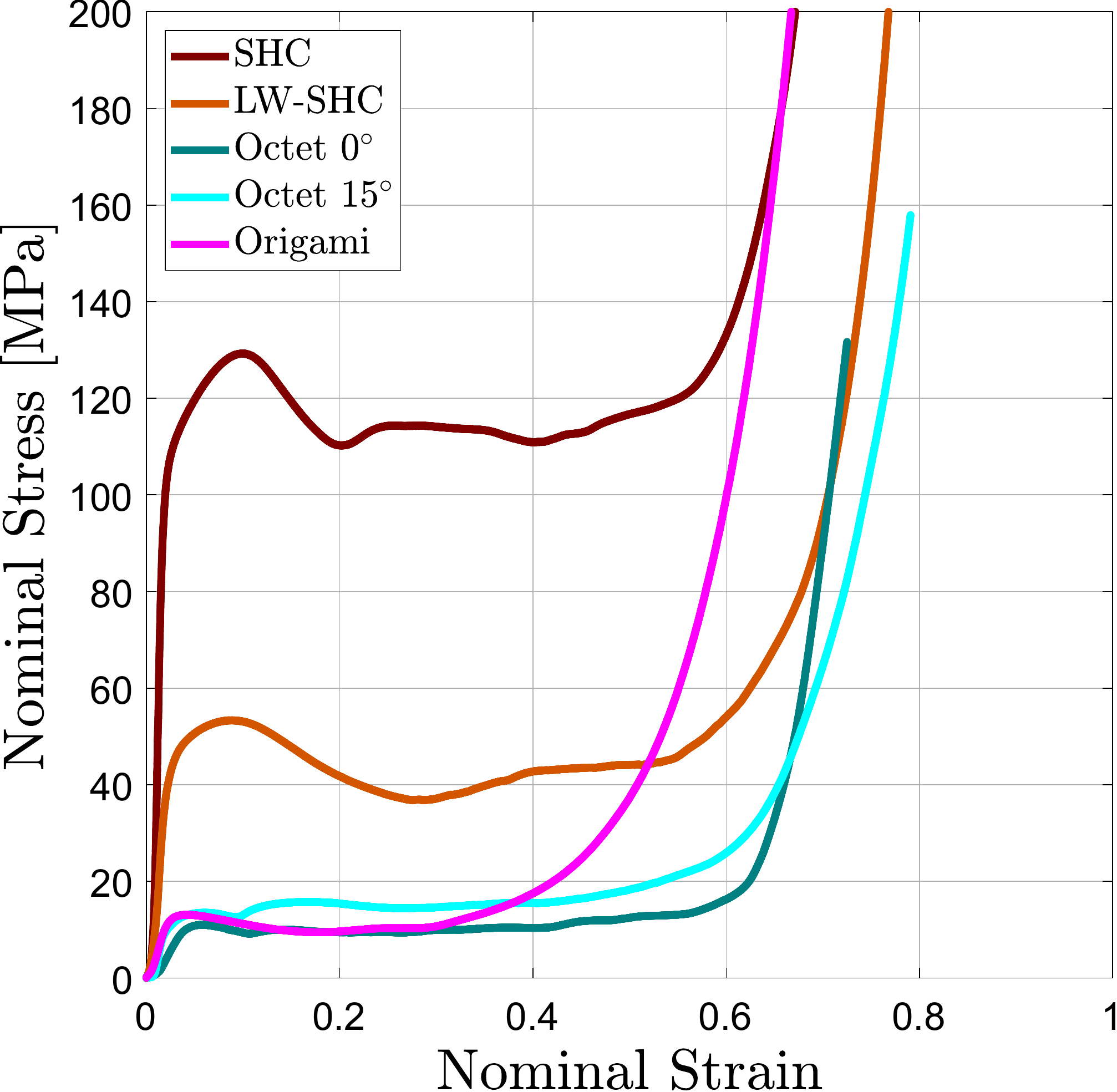}
		\caption{}
		\label{Fig:125WNomNom2}
	\end{subfigure}
	\caption{Nominal stress versus strain of the five different geometries manufactured with process variant 2: (a) dynamically tested at an impact speed of 100 m/s and (b) tested quasi-statically.}
	\label{Fig:NominalStressNominalStrain5geometriesDYNandQS}
\end{figure}

The nominal stress versus strain curves from the impact tests are shown in Fig. \ref{Fig:NominalStressNominalStrain5geometriesDYNandQS}(a). For comparison, the equivalent quasi-static curves are plotted in Fig. \ref{Fig:NominalStressNominalStrain5geometriesDYNandQS}(b). The quasi-static and dynamic curves have a similar shape, noting that the instrumentation rise time will affect the accuracy of the initial stress rise in the dynamic case. There is an increase in the densification strain under dynamic loading, particularly for the lower strength geometries. For the lattice structures this may be due to the delayed buckling of the struts observed in the high speed photography. For the origami structure, the non-uniform collapse and reduced auxetic effect under dynamic loading may lead to closer packing of the cell walls.

\begin{figure}[H]
	\centering
	\includegraphics[scale=0.6]{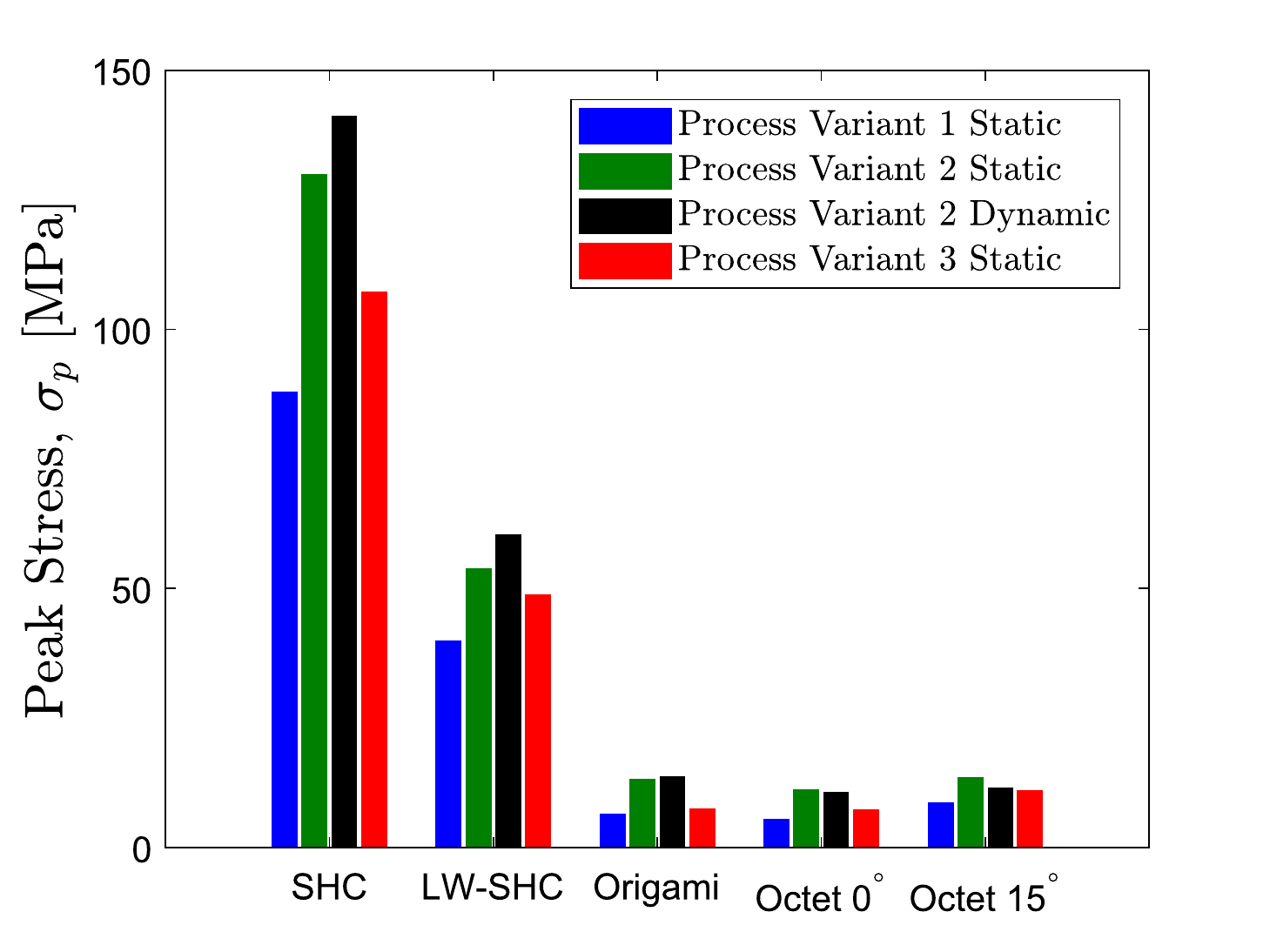}
	\caption{Comparison of the peak stresses of the five geometries manufactured with the three process variants and tested quasi-statically and dynamically. Note that in the dynamic tests the stress is measured at the supported face.}
	\label{Fig:PeakStressesCellularQSvsDyn}
\end{figure}

A comparison of the peak stresses measured in the quasi-static and dynamic experiments is given in Fig. \ref{Fig:PeakStressesCellularQSvsDyn}. A dynamic elevation in peak strength is observed for the two highest strength geometries, the square honeycomb and the lattice-walled honeycomb. This can be attributed to the effect of inertia on buckling \cite{Radford2007,Harris2017}. However, the effect is relatively small, and the trends in peak strength across the different cellular geometries are very similar for the quasi-static and dynamic load cases.

Dynamic loading with a 100 m/s impact speed therefore leads to some changes in the collapse behaviour of the cellular materials, particularly those of lower strength. However, the essential features of the compressive response are similar to the quasi-static case. We therefore conclude that the insights drawn from the quasi-static tests on the energy absorbing performance of these LPBF cellular materials are still applicable up to the impact speeds considered here.

\section{Conclusions}
\label{Sec:Conclusions}

In this work, we measure the energy absorbing properties of 316L stainless steel LPBF cellular materials to assess the sensitivity to changes in the LPBF processing parameters. Both strut-based and facet-based architectures are experimentally investigated. Three process parameter sets are considered, using laser powers of 50 W, 125 W and 200 W (denoted variants 1, 2 and 3), with the exposure time adjusted to maintain a fixed total heat input. Our main findings are:
\begin{enumerate}
	\item The intermediate power and exposure parameter set (variant 2) shows negligible porosity, with porosity increasing for lower power and longer exposure (variant 1), and for higher power and shorter exposure (variant 3).
	\item The choice of process parameters and the resulting microstructures influenced the density, yield strength, ductility, hardness and energy absorption measured, with variant 2 showing the best mechanical performance. This was observed for both tensile dogbone and cellular specimens. In the dogbone samples, porosity was found to reduce the yield strength by a factor of $\left(1 - p_J \right)^2$, where $p_J$ is the porosity. 
	\item The choice of cellular architecture had a greater influence on mechanical performance than the choice of process parameters. In terms of strength and energy absorption, the material property space occupied by each cellular geometry remained distinct when the process parameters were moved away from process parameter set 2.
	\item Dynamic impact tests up to 100 m/s revealed a modest increase in peak strength for the higher strength specimens, due to inertial effects. A change in the collapse mode was also seen for the lower strength specimens in particular, including deformation localisation (for the origami structure) and delayed buckling (for the octet trusses). However, the key features of the compressive stress versus strain response were similar for the dynamic (distal face) and quasi-static load cases. The quasi-static measurements, therefore, provide relevant insights into the LPBF cellular material process-property interactions for applications in impact energy absorption.
\end{enumerate}

\section{Acknowledgments}
\label{Sec:Acknowledge of funding}

\noindent M. Simoes acknowledges financial support from the Engineering and Physical Sciences Research Council (grant EP/R512461/1) and from the European Space Agency (grant NPI 565 - 2017).

\small 

\bibliographystyle{elsarticle-num}
\bibliography{library}

\end{document}